\documentclass[twocolumn, aps, prd, eqsecnum, showpacs,
  preprintnumbers, superscriptaddress]{revtex4}
\usepackage{graphicx, amsmath}
\def\aa{Astron.\ Astrophys.\ }
\def\apjs{Astrophys.\ J.\ Suppl.\ }
\def\cqg{Class.\ Quantum\ Grav.\ }
\def\grg{Gen.\ Relativ.\ Grav.\ }
\def\jcp{J.\ Compt.\ Phys.\ }
\def\mnras{Mon.\ Not.\ R.\ Astron.\ Soc.\ }
\def\prsla{Proc.\ R.\ Soc.\ London\ A\ }
\begin{document}
\title
{Collapse of differentially rotating supermassive stars: Post black
  hole formation}
%
\author{Motoyuki Saijo}
\email{saijo@rikkyo.ac.jp}
%
\affiliation
{Department of Physics, Rikkyo University,
Toshima, Tokyo 171-8501, Japan}
\affiliation
{Research Center for Measurement in Advance Science, Rikkyo University,
Toshima, Tokyo 171-8501, Japan}
%
\author{Ian Hawke}
\email{I.Hawke@soton.ac.uk}
%
\affiliation
{School of Mathematics, University of Southampton, 
Southampton SO17 1BJ, United Kingdom}
%
\received{27 May 2009}
\accepted{13 August 2009}
%
\begin{abstract}
  We investigate the collapse of differentially rotating supermassive
  stars (SMSs) by means of 3+1 hydrodynamic simulations in general
  relativity.  We particularly focus on the onset of collapse to
  understand the final outcome of collapsing SMSs.  We find that the
  estimated ratio of the mass between the black hole (BH) and the
  surrounding disk from the equilibrium star is roughly the same as
  the results from numerical simulation.  This suggests that the
  picture of axisymmetric collapse is adequate, in the absence of
  nonaxisymmetric instabilities, to illustrate the final state of the
  collapse.  We also find that quasi-periodic gravitational waves
  continue to be emitted after the quasinormal mode frequency has
  decayed.  We furthermore have found that when the newly formed BH is
  almost extreme Kerr, the amplitude of the quasi-periodic oscillation
  is enhanced during the late stages of the evolution.  Geometrical
  features, shock waves, and instabilities of the fluid are
  suggested as a cause of this amplification behaviour.  This
  alternative scenario for the collapse of differentially rotating
  SMSs might be observable by LISA.
\end{abstract}
%
\maketitle
%
\section{Introduction}
\label{sec:intro}

There exists plenty of evidence that supermassive black holes (SMBHs)
exist in the centre of galaxies, but their actual formation process
has been a mystery for many decades \citep{Rees03}.  Several different 
scenarios have been proposed, some based on stellar dynamics, others
on gas hydrodynamics, and still others that combine these processes.
There are three major routes to form a SMBH.  One possibility is
that the gas in the newly-forming galaxy does not break up into
stellar-mass condensations, but instead forms a single superstar, and
then undergoes complete gravitational collapse.  Another possibility
is that instead of forming a single superstar, black holes (BHs) might
grow from smaller seeds, and then merge to form a SMBH.  The final
possibility is the runaway growth of a stellar mass BH to a SMBH by
accretion.  Here we consider the possibility of forming a SMBH from
the collapse of a supermassive star (SMS).

There are two categories of collapsing rotating SMSs based on their
angular momentum distribution.  One is the collapse of a uniformly
rotating SMS.  This happens when momentum transport is large, either
through viscous turbulence or magnetic processes, which drives the star
to rotate uniformly.  In this case the path to the SMBH is the
following.  First the SMS contracts until the mass shedding limit is
reached, conserving the angular momentum of the star.  If the SMS
contains sufficient angular momentum, the star evolves
quasi-stationary along the mass-shedding limit, releasing the mass and
angular momentum.  When the star reaches the post-Newtonian
instability it begins to collapse.  \citet{BS99} found that the star
starts to collapse from the universal configuration when the system
contains sufficient angular momentum, independent of the history of
the star.  \citet{SBSS02} found that the collapse is coherent.  That
is, no significant bar instability occurs before BH formation in three 
dimensional post-Newtonian simulations.  \citet{SS02} found that a
final Kerr BH of $a/M \approx 0.75$ contains 90\% of the total
rest-mass of the system, and that the disk around the BH contains 10\%
of that in an axisymmetric general relativistic simulation. 

The other category of collapsing rotating SMSs are differentially
rotating.  This happens when the viscous and the magnetic effects are
small, which allows the star to rotate differentially.  One of the
representative scenarios for forming a differentially rotating star is
as follows.  First, a gas cloud gathers in an almost spherical
configuration with some amount of angular momentum in the system.  Next
the almost spherical star contracts, conserving specific angular
momentum due to the lack of viscosity, to form a differentially
rotating star, and possibly a disk at the end of the contraction
\citep{BO73}.
   
During the contraction of the differentially rotating SMS, prior to
forming a supermassive disk, two possible instabilities may arise that
terminate the contraction.  One is the post-Newtonian gravitational
instability, which leads the star to collapse dynamically.  The other
is the dynamical bar mode instability, which changes the angular
momentum distribution of the star to form a bar, and possibly leads to
the central core of the star collapsing to a BH due to the angular
momentum loss.

One of the primary observational missions for space-based detection of
gravitational waves is the investigation of supermassive objects
\citep{Thorne98}.  Since the Laser Interferometer Space Antenna (LISA)
will have long arms ($10^{6} \, {\rm km}$), the detector will be most
sensitive in the low frequency band ($10^{-4} \sim 10^{-1}$ Hz).
Potential sources of high signal to noise events in this frequency
range include quasi-periodic waves arising from nonaxisymmetric bars
in collapsing SMSs and the inspiral of binary SMBHs
(e.g.\ Ref.~\citep{SS09}).  In addition, the nonspherical collapse of
rotating SMSs to SMBHs could be a significant source of burst and
quasi-normal gravitational waves (e.g.\ Ref.~\citep{FN03}).  In this
paper we track the collapse of a SMS by numerical simulation to
investigate some of these possibilities.

\begin{figure}
\centering
\includegraphics[keepaspectratio=true,width=8cm]{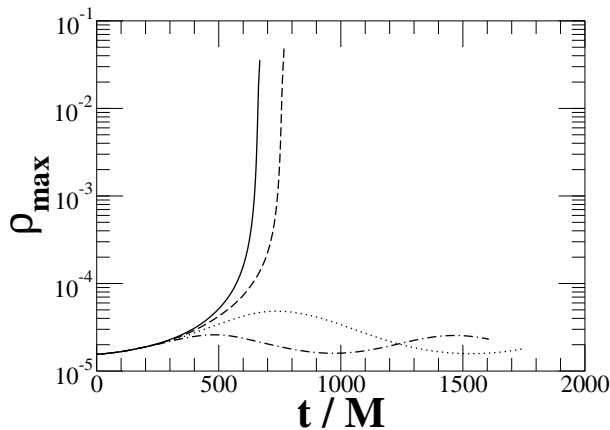}
\caption{
Maximum of the rest mass density as a function of time.  Solid,
dashed, dotted, and dash-dotted lines denote models I, II, III, IV,
respectively.
}
\label{fig:collapse}
\end{figure}

Here we focus on the post-Newtonian gravitational instability in
differentially rotating SMSs.  The aim of this work is to understand
the final fate of the collapse of differentially rotating SMSs.  There
are two studies that address this issue.  \citet{NS01} studied the
``zero-viscosity'' equilibrium sequence in differentially rotating
SMSs in Newtonian gravity.  They showed that the star reaches the
bifurcation point of dynamical bar instability before the disk forms
by contraction.  \citet{Saijo04} studied the gravitational collapse of
differentially rotating SMSs in the conformally flat approximation of
general relativity near the onset of radial instability.  He showed
that once the star collapses, the collapse is coherent and may form a
SMBH with no substantial nonaxisymmetric deformation due to
instability.  However the previous work may contain less angular
momentum so that there might be no significant difference between the
cases of uniform rotation and differential rotation.  Therefore we now
focus on the case where the angular momentum is more contained in the
star than the previous case, which potentially leads to rotational
instabilities if they occur.  In particular, we plan to answer the
following questions.  Does the BH form coherently?  What are the
features of the dynamics?  Does the newly formed disk lead to
instabilities?  Can this system act as an efficient source of
gravitational waves (GWs)?  In order to answer these questions, three
dimensional general relativistic hydrodynamics are desirable.

The content of this paper is as follows.  In Sec.~\ref{sec:bequation},
we briefly explain the general relativistic hydrodynamics, especially
the numerical tools we used to simulate the violent phenomena.  In
Sec.~\ref{sec:nr}, we introduce our findings of dynamical BHs, disk
formation, and gravitational waves.  Section~\ref{sec:Conclusion} is
devoted to the summary of this paper.  Throughout this paper, we use
the geometrized units with $G=c=1$ and adopt Cartesian coordinates
$(x,y,z)$ with the coordinate time $t$.  Note that Greek index takes
$(t,x,y,z)$, while Latin one takes $(x,y,z)$.

\section{Basic equations}
\label{sec:bequation}
In this section, we briefly describe the three-dimensional
relativistic hydrodynamics in full general relativity.

\subsection{The gravitational field equations}

We define the spatial projection tensor $h^{\mu\nu} \equiv g^{\mu\nu}
+ n^{\mu} n^{\nu}$, where $g^{\mu\nu}$ is the spacetime metric,
$n^{\mu} = (1/\alpha, -\beta^i/\alpha)$ the unit normal to a spatial
hypersurface, and where $\alpha$ and $\beta^i$ are the lapse and
shift.  Firstly, we introduce a conformal factor $\psi = e^{\phi}$ to
the spatial three metric $\gamma_{ij}$ as
\begin{equation}
\tilde{\gamma}_{ij} = e^{-4\phi} \gamma_{ij},
\end{equation}
and choose the determinant of the conformally related metric
$\tilde{\gamma}_{ij}$ to be unity, $\phi = (\ln \gamma) / 12$.
Secondly, we introduce a trace-free part of the extrinsic curvature
$A_{ij}$, and its conformally related tensor $\tilde{A}_{ij}$, whose
indices will be raised and lowered by the conformally related metric
$\tilde{\gamma}_{ij}$, as in Ref.~\citep{SN95},
\begin{equation}
\tilde{A}_{ij} = e^{-4 \phi} A_{ij} = 
  e^{-4 \phi} \left( K_{ij} - \frac{1}{3} \gamma_{ij} K \right),
\end{equation}
where $K_{ij}$ is the extrinsic curvature.  Finally we introduce the
conformal connection functions $\tilde{\Gamma}^{i}$ \citep{BS98}
\begin{equation}
\tilde{\Gamma}^i \equiv \tilde{\gamma}^{jk} \tilde{\Gamma}^{i}_{jk} =
- \frac{\partial}{\partial x^{j}} \tilde{\gamma}^{ij},
\end{equation}
as a variable to describe the Ricci tensor of the conformally related
metric.  This step, the expansion of the Ricci tensor in terms of the
metric based variables, plays a crucial role in maintaining a long time
stable evolution of Einstein's field equations \citep{SN95}.

\begin{table*}[htbp]
\begin{center}
\caption{
Four different rotating equilibrium SMSs for evolution}
\begin{ruledtabular}
\begin{tabular}{c c c c c c c c c c}
Model & spacetime &
$R_{\rm p} / R_{\rm e}$\footnotemark[1] & 
$\rho_{0}^{\rm max}$\footnotemark[2] & 
$M$\footnotemark[3] &
$T/W$\footnotemark[4] &
$J/M^2$\footnotemark[5] & 
$M/R$\footnotemark[6] & 
$m_{\rm disk}$\footnotemark[7] & 
$(a/M)^{\rm (BH)}$\footnotemark[8]
\\
\hline
  I & GR\footnotemark[9] & $0.450$ & $1.56 \times 10^{-5}$ & $4.88$ &
  $0.108$ & $0.99$ & $2.56 \times 10^{-2}$ & --- & ---
\\
  I & CF\footnotemark[10] & $0.450$ & $1.56 \times 10^{-5}$ & $4.88$ &
  $0.108$ & $0.99$ & $2.56 \times 10^{-2}$ & $0.044$ & $0.98$
\\
 II & GR & $0.425$ & $1.56 \times 10^{-5}$ & $5.07$ & $0.118$ & $1.03$
 & $2.63 \times 10^{-2}$ & --- & ---
\\
 II & CF & $0.425$ & $1.56 \times 10^{-5}$ & $5.07$ & $0.118$ & $1.03$
 & $2.63 \times 10^{-2}$ & --- & $\gtrsim 1$
\\
III & GR & $0.400$ & $1.56 \times 10^{-5}$ & $5.31$ & $0.131$ & $1.07$
& $2.78 \times 10^{-2}$ & --- & ---
\\
 IV & GR & $0.375$ & $1.56 \times 10^{-5}$ & $5.75$ & $0.156$ &
 $1.10$ & $3.47 \times 10^{-2}$ & --- & ---
\\
\end{tabular}
\end{ruledtabular}
\label{tab:equilibrium}
\footnotetext[1]{Ratio of the polar proper radius to the equatorial
  proper radius}
\footnotetext[2]{Maximum rest mass density}
\footnotetext[3]{Gravitational mass}
\footnotetext[4]{Ratio of the rotational kinetic energy to the
  gravitational binding energy}
\footnotetext[5]{$J$: Total angular momentum}
\footnotetext[6]{$R$: Circumferential radius}
\footnotetext[7]{Ratio of the estimated rest mass of the disk to the 
  rest mass of the equilibrium star}
\footnotetext[8]{Estimated Kerr parameter of the final hole}
\footnotetext[9]{General theory of relativity}
\footnotetext[10]{Conformally flat spacetime}
\end{center}
\end{table*}

Therefore we evolve the spacetime with the 17 spacetime associated
variables ($\phi$, $K$, $\tilde{\gamma}_{ij}$, $\tilde{A}_{ij}$,
$\tilde{\Gamma}^{i}$) as
\begin{widetext}
\begin{eqnarray}
\left( \frac{\partial}{\partial t} - {\cal L}_{\beta} \right) 
\phi &=& 
-\frac{1}{6} \alpha K, 
\label{eqn:evolution_phi}
\\
\left( \frac{\partial}{\partial t} - {\cal L}_{\beta} \right) 
K &=& 
- \gamma^{ij} D_{i} D_{j} \alpha 
+ \alpha \left[ \tilde{A}_{ij} \tilde{A}^{ij} +\frac{1}{3} K^{2} +
  \frac{1}{2} (\rho_{\rm H} + S) \right],
\\
\left( \frac{\partial}{\partial t} - {\cal L}_{\beta} \right) 
\tilde{\gamma}_{ij} &=&
- 2 \alpha \tilde{A}_{ij},
\\
\left( \frac{\partial}{\partial t} - {\cal L}_{\beta} \right) 
\tilde{A}_{ij} &=&
e^{-4\phi} [ -D_{i} D_{j} \alpha + \alpha (R_{ij} - S_{ij}) ]^{\rm TF}
+ \alpha ( K \tilde{A}_{ij} - 2 \tilde{A}_{il} \tilde{A}^{l}_{j} ),
\\
\left( \frac{\partial}{\partial t} - {\cal L}_{\beta} \right) 
\tilde{\Gamma}^i &=&
- 2 \tilde{A}^{ij} \frac{\partial}{\partial x^{j}} \alpha + 2 \alpha 
\left( 
  \tilde{\Gamma}^{i}_{jk} \tilde{A}^{jk} 
  - \frac{2}{3} \tilde{\gamma}^{ij} \frac{\partial}{\partial x^{j}} K 
  - \tilde{\gamma}^{ij} S_{j} 
  + 6 \tilde{A}^{ij} \frac{\partial}{\partial x^j} \phi
\right)
\nonumber \\
&&
- \frac{\partial}{\partial x^j}
\left(
  \beta^{l} \frac{\partial}{\partial x^l} \tilde{\gamma}^{ij} 
  - 2 \tilde{\gamma}^{m (j} \frac{\partial}{\partial x^{m}} \beta^{i)} 
  + \frac{2}{3} \tilde{\gamma}^{ij} \frac{\partial}{\partial x^{l}}
  \beta^{l}
\right),
\end{eqnarray}
\end{widetext}
where ${\cal L}_{\beta}$ denotes the Lie derivative along the shift
$\beta^{i}$, $\rho_{\rm H} = T_{\mu \nu} n^{\mu} n^{\nu}$, $S_i =
T_{\mu \nu} n^{\mu} h^{\nu}_{i}$ and TF the trace-free part of the
tensor.  This set of equations when used to solve Einstein's field
equations numerically is usually called the BSSN formalism. 

We use as a slicing condition the generalised hyperbolic $K$-driver
\citep{ABDKPST03,PRRSADDDKNS07}
\begin{equation}
\left( 
  \frac{\partial}{\partial t} - \beta^i \frac{\partial}{\partial x^i}
\right)
\alpha = - \alpha^2 f(\alpha) (K - K_0),
\end{equation}
where $f(\alpha)=2/\alpha$, $K_0=K_{t=0}$ in order to avoid
singularities.  As for the shift, we use the generalised hyperbolic
$\tilde{\Gamma}$-driver \citep{BCCKM06}
\begin{eqnarray}
\left( 
  \frac{\partial}{\partial t} - \beta^j \frac{\partial}{\partial x^j}
\right)
\beta^i &=& \frac{3}{4} \alpha B^{i}
,\\
\left( 
  \frac{\partial}{\partial t} - \beta^j \frac{\partial}{\partial x^j}
\right)
B^i &=&
\left( 
  \frac{\partial}{\partial t} - \beta^j \frac{\partial}{\partial x^j}
\right)
\tilde{\Gamma}^i - \eta B^i
,\nonumber \\
&&
\label{eqn:gauge_B}
\end{eqnarray}
where $\eta$ is a damping parameter.

\begin{figure}
\centering
\includegraphics[keepaspectratio=true,width=8cm]{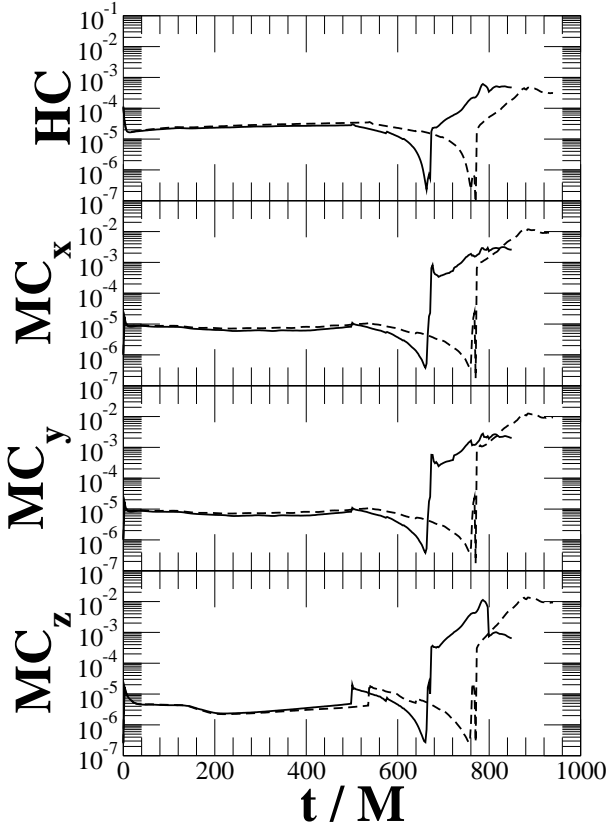}
\caption{
Hamiltonian constraint (solid line) and Momentum constraints (dashed
lines) for models I and II throughout the evolution.  Note that we
normalised $HC$ with $16 \pi \rho_{\rm max}$ and $MC_i$ with $8 \pi
\rho_{\rm max} \sqrt{M/R}$, where $M$ is the gravitational mass, $R$
the circumferential radius of the equilibrium star, respectively.
}
\label{fig:const}
\end{figure}

\subsection{The matter equations}
For a perfect fluid, the energy momentum tensor takes the form 
\begin{equation}
T^{\mu \nu} = \rho 
  \left( 1 + \varepsilon + \frac{P}{\rho} \right) u^{\mu} u^{\nu} 
  + Pg^{\mu\nu},
\end{equation}
where $\rho$ is the rest-mass density, $\varepsilon$ the specific
internal energy, $P$ the pressure, $u^{\mu}$ the four-velocity.

We adopt a $\Gamma$-law equation of state in the form
\begin{equation}
P = (\Gamma - 1) \rho \varepsilon,
\label{gammalaw1}
\end{equation}
where $\Gamma$ is the adiabatic index which we set to $4/3$ in this
paper, representing a SMS (the pressure is dominated by radiation).

In the absence of thermal dissipation, Eq.~(\ref{gammalaw1}), together
with the first law of thermodynamics, implies a polytropic equation of
state
\begin{equation}
P = \kappa \rho^{1+1/n}, 
\label{gammalaw}
\end{equation}
where $n=1/(\Gamma-1)$ is the polytropic index and $\kappa$ a
constant.

From $\nabla_{\mu} T^{\mu\nu}=0$ together with the continuity
equation, the flux conservative form of the relativistic continuity,
the relativistic energy and the relativistic Euler equations are
\citep{BFIMM97}
\begin{equation}
\frac{1}{\sqrt{-g}} \frac{\partial}{\partial t} 
  (\sqrt{\gamma} {\boldsymbol{\cal U}}) + 
\frac{1}{\sqrt{-g}} \frac{\partial}{\partial x^i} 
  (\sqrt{-g} {\boldsymbol{\cal F}}^i)
= \boldsymbol{\cal S}^i,
\end{equation}
where the state vector $\boldsymbol{\cal U}$, the flux vectors
$\boldsymbol{\cal F}^i$, and the source vectors $\boldsymbol{\cal
  S}^i$ are
\begin{widetext}
\begin{eqnarray}
\boldsymbol{\cal U} &=& 
  [D, S_i, \tau]^{T},\\
\boldsymbol{\cal F}^i &=&
  \left[ D \left( v^i - \frac{\beta^i}{\alpha} \right), 
   S_j \left( v^i - \frac{\beta^i}{\alpha} \right) + P \delta^{i}_{j},
   \tau \left( v^i - \frac{\beta^i}{\alpha} \right) + P v^{i}
  \right]^{T},\\
\boldsymbol{\cal S}^i &=&
  \left[0, 
    T^{\mu \nu} 
    \left( 
      \frac{\partial}{\partial x^{\mu}} g_{\nu j} -
      \Gamma^{\delta}_{\nu \mu} g_{\delta j}
    \right),
    \alpha 
    \left( 
      T^{\mu 0} \frac{\partial}{\partial x^{\mu}} \ln \alpha -
      T^{\mu \nu} \Gamma^{0}_{\nu \mu}
    \right)
  \right]^{T}.
\end{eqnarray}
\end{widetext}
Note that $(\rho, v_i, \varepsilon)$ are the physical variables of the
above equations, and the conserved quantities $D$, $S_i$, $\tau$ are
\begin{eqnarray}
D &=& \rho W, \\
W &=& \alpha u^t, \\
S_i &=& \rho h W^2 v_i, \\ 
E &=& \rho h W^2 - P,\\
\tau &=& E - D,
\end{eqnarray}
where $v_{j} = u_{j} / W$, $h \equiv 1 + \varepsilon + P / \rho$ is
the specific enthalpy.  In the Newtonian limit, the above three
physical variables coincide to the rest mass density, the flux density
of the rest mass, and the energy of a unit volume of fluid.

\begin{figure}[t]
\centering
\includegraphics[keepaspectratio=true,width=8cm]{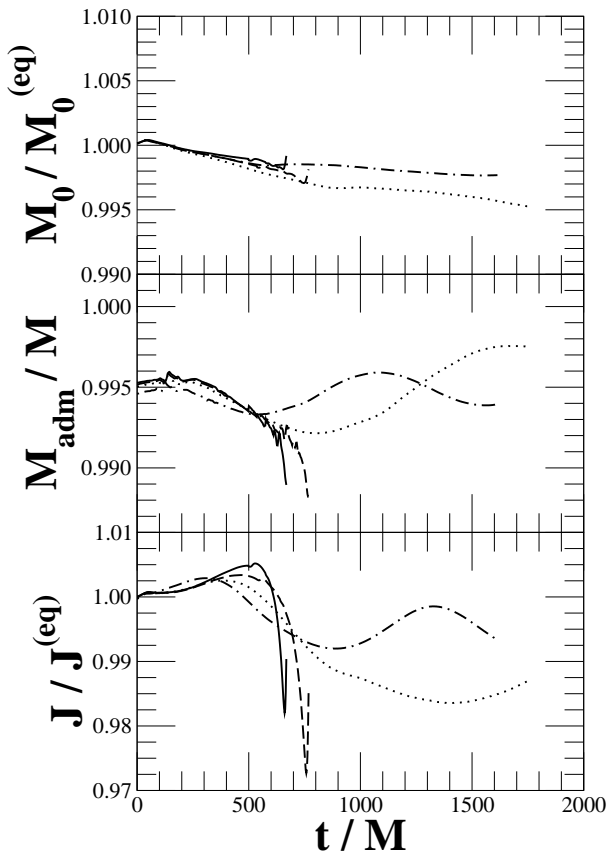}
\caption{
Rest mass, ADM mass and the total angular momentum throughout the
evolution.  We do not plot the above three quantities after the
apparent horizon formation.  Solid, dashed, dotted and dash-dotted
line denote model I, II, III and IV, respectively.  Since we use a
different computational domain for computing a equilibrium and
evolution quantities, the relative amount rate of $\approx 0.5\%$
appears in the ADM mass.
}
\label{fig:raj}
\end{figure}

\begin{figure}
\centering
\includegraphics[keepaspectratio=true,width=8cm]{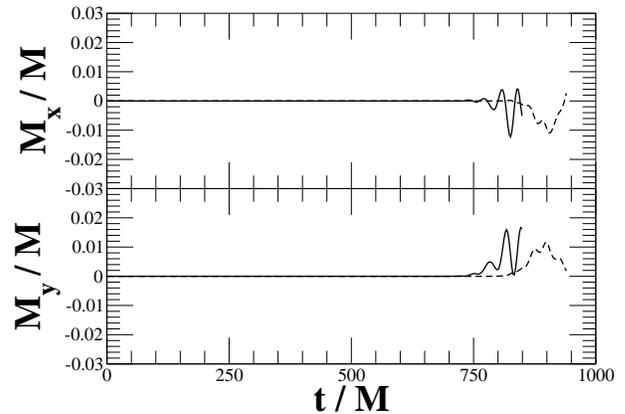}
\caption{
Centre of mass throughout the evolution.  Solid and dash line denotes
models I and II, respectively.
}
\label{fig:com}
\end{figure}

\subsection{Dynamical horizon}
\label{sec:DynamicalHorizon}
A dynamical horizon is defined as the spacelike marginally trapped
tube which is composed of future-marginally trapped surfaces on which
the expansion associated with the ingoing null direction is strictly
negative \citep{AK03}. If it is the outermost trapped surface, which
is expected to be the case with all those considered here, then it is
an apparent horizon. An angular momentum of the horizon $J_{\rm BH}$
can be written (see e.g.\ section III.B of Ref.~\citep{DH06}) as 
\begin{equation}
J_{\rm BH} = - \frac{1}{8 \pi} \int_{S_{\rm R}} K_{\mu \nu} R^{\mu}
\varphi^{\nu} ds,
\end{equation}
where $R^{\mu}$ is the outward directed spacelike normal to the
horizon within the spacelike slice, and $\varphi^{a}$ is a rotational
vector field on the horizon. This can be interpreted as the Komar
angular momentum when $\varphi^{a}$ is a rotational Killing vector on
the horizon. The code and method used to compute numerically an
approximate Killing vector, should it exist, is described in
Ref.~\citep{DH03}.

The dynamical horizon mass of the hole $M_{\rm BH}$ can be computed
once we have the angular momentum of the hole as
\begin{equation}
M_{\rm BH} = \frac{1}{2 R_{\rm s}}\sqrt{R_{\rm s} + 4 J_{\rm BH}^{2}}.
\end{equation}
Note that $R_{\rm s}$ is the area radius of the hole.  We also monitor
the dimensionless Kerr parameter $(J / M^2 )_{\rm BH}$ throughout the
evolution.

\subsection{Gravitational waveform}
In order to monitor gravitational waves, we use the
Newman-Penrose-Weyl scalar $\Psi_4$
\begin{equation}
\Psi_{4} = 
  C_{\mu \nu \lambda \sigma} k^{\mu} \bar{m}^{\nu} k^{\lambda} \bar{m}^{\sigma},
\end{equation}
where $C_{\mu \nu \lambda \sigma}$ is the Weyl tensor, $k^{\mu}$ the
ingoing null vector, $m^{\mu}$ and $\bar{m}^{\mu}$ are the orthogonal
spatial-null vectors of the four complex null tetrad ($l^{\mu}$,
$k^{\mu}$, $m^{\mu}$, $\bar{m}^{\mu}$).  More precisely, a set of null
vectors are defined by introducing a orthonormal tetrad
$e^{\mu}_{(a)}$ as
\begin{eqnarray}
l^{\mu} &=& 
  \frac{1}{\sqrt{2}} \left( e_{(0)}^{\mu} + e_{(1)}^{\mu} \right)
,\\
k^{\mu} &=& 
  \frac{1}{\sqrt{2}} \left( e_{(0)}^{\mu} - e_{(1)}^{\mu} \right)
,\\
m^{\mu} &=& 
  \frac{1}{\sqrt{2}} \left( e_{(2)}^{\mu} + i e_{(3)}^{\mu} \right)
,\\
\bar{m}^{\mu} &=& 
  \frac{1}{\sqrt{2}} \left( e_{(2)}^{\mu} - i e_{(3)}^{\mu} \right)
.
\end{eqnarray}
Note that we choose the vector $e^{\mu}_{(0)}$ as the unit normal to
the spatial hypersurface, $e^{\mu}_{(1)}$ the unit radial vector in
spherical coordinates, and ($e^{\mu}_{(2)}$, $e^{\mu}_{(3)}$) the unit
vectors in the angular directions.  The Weyl scalar $\Psi_4$
represents the outgoing gravitational waves at infinity
\begin{equation}
\Psi_{4} = \ddot{h}_{+} - i \ddot{h}_{\times},
\end{equation}
where $h_{+}$ and $h_{\times}$ are the two polarization modes
(transverse-traceless condition) of the perturbed metric from flat
spacetime in spherical coordinate and $\dot{q}$ represents the time
derivative of the quantity $q$.  If we put the observer sufficiently
far from the source, the Weyl scalar $\Psi_4$ roughly represents the
outgoing gravitational waves, ignoring the radiation back scattered by
the curvature.  Therefore we monitor the Weyl scalar $\Psi_4$ to
understand key features of the gravitational waves emitted by this
system.

\subsection{Treatment of the spacetime singularity from the newly
  formed black hole}
When a horizon is found in the simulation some of the matter within
the horizon is excised from the domain using the technique of
Ref.~\citep{Hawke05}. The domain excised is a sphere with coordinate
radius half the minimum coordinate radius of the horizon at the time
of excision. It is allowed to grow during evolution but not to shrink
(in coordinate radius).  Although the spacetime is not excised, we
introduce a strong artificial dissipation term in
Eqs.~(\ref{eqn:evolution_phi}) -- (\ref{eqn:gauge_B}), combined with
the standard gauge conditions, to avoid problems with the singularity
\citep{BR06}.  In practice, we add a dissipation term $- \epsilon
dx^{i} \partial_{i}^{4} q$ in Eqs.~(\ref{eqn:evolution_phi}) --
(\ref{eqn:gauge_B}), where $q$ is the quantity evolved by the
particular evolution equation.  We specify the dissipation parameter
$\epsilon \approx 0.2$ for our simulation.

In addition to this dissipation term, we have found it necessary to
stabilize our evolutions in certain circumstances by reducing the
timestep on individual refinement levels. This is particularly true
for the finest levels near the horizon where low density matter is
orbiting at extreme speed.

\begin{figure*}
\centering
\includegraphics[keepaspectratio=true,width=14cm]{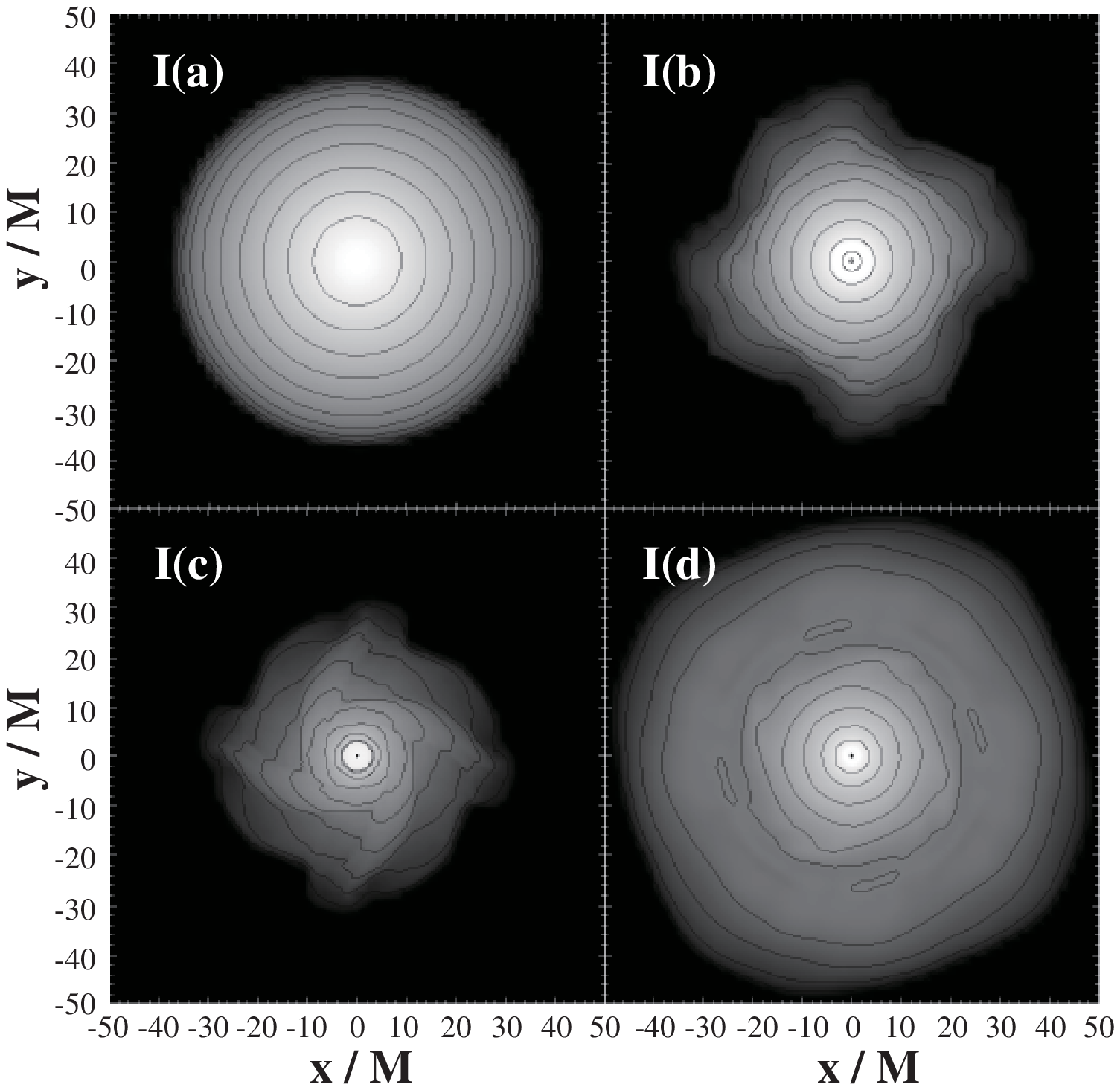}
\caption{
Contours of the rest mass density in the equatorial plane for model
I.  Snapshots are plotted at values of ($t$, $\rho_{\rm max}$) = (a)
($0$, $1.56 \times 10^{-5}$), (b) ($671 M$, $5.04 \times 10^{-2}$),
(c) ($755 M$, $5.28 \times 10^{-5}$) and (d)  ($839 M$, $1.68 \times
10^{-5}$), with a cutoff density of $\rho_{\rm cut} = 1.56 \times
10^{-13}$, respectively.  The contour lines denote rest mass densities
$\rho/\rho_{\rm max} = 10^{-(15-i)d}$ ($i=1, \cdots, 14$), where
$d=(\log \rho_{\rm max} - \log \rho_{\rm cut})/15$.  Note that the
apparent horizon exists after $t = 671 M$, and the coordinate radius
of the apparent horizon in the equatorial plane are $r_{\rm hrz} =$
(c) $0.283 M$ and (d) $0.417 M$, respectively.
}
\label{fig:qxy1}
\end{figure*}

\subsection{Computational codes and techniques}
We perform 3+1 hydrodynamic simulations in general relativity using
{\sc CACTUS} \citep{Cactus} (gravitational physics), {\sc CARPET}
\citep{Carpet} (mesh refinement of space and time), {\sc WHISKY} 
\citep{Whisky1} (see Ref.~\citep{WhiskyAccurate} and references
therein for a recent description) (general relativistic
hydrodynamics).  We set the outermost boundary of the computational 
grid for all direction as $x_{\rm max} = 111$ -- $131 M$, imposing
plane symmetry across the $z=0$ plane, and use 5 -- 11 refinement
levels to maintain the resolution at the central core region.   We set
the $i$-th refinement level of the uniform grid as $L_{i} = 2^{7-i}
\times 10$, $\Delta r_i = 2^{5-i}$ ($i = 1, \cdots , 11$ for
collapsing objects), where $L_i$ is the outer boundary of the grid,
$\Delta r_i$ is the stepsize of the spatial grid, respectively.  The
finest central resolution for the collapsing SMS is $\Delta x = (3.08$ 
-- $3.21) \times 10^{-3} M$.  The Courant condition is set as $\Delta
t = C_{\rm crt} \Delta x$, where $C_{\rm crt}= 0.05$, $0.1$, $0.2$ for
the finest three refinement levels for collapsing stars and $C_{\rm
  crt} = 0.4$ for the rest of the refinement levels, damping parameter
of the shift condition $\eta = 3.91$ -- $4.05$ $M^{-1}$ for the
collapsing stars.  As in previous works we use a
reconstruction-evolution method to evolve the matter, with PPM
reconstruction~\citep{Collela84} of the matter variables and
Marquina's formula~\citep{AIM99} to compute the inter-cell
fluxes.  Third order Runge-Kutta evolution in time is also used in our 
computation.

\section{Numerical Results}
\label{sec:nr}
\subsection{Equilibrium Stars}
We construct the four differentially rotating SMSs for evolution.  We
use the polytropic equation of state of Eq.~(\ref{gammalaw}) with
$n=3$, which represents the radiation pressure dominant, SMS sequence,
and the relativistic ``$j$-constant'' rotation law as
\begin{equation}
u^{t} u_{\varphi} = A^{2} (\Omega_{c} - \Omega),
\end{equation}
where $A (\equiv R_{e}\hat{A} )$ represents the degree of differential 
rotation and has the dimension of length, $\Omega_{c}$ the central
angular velocity, and $\Omega$ the angular velocity.  We choose
$\hat{A}=0.3$ here.  In the Newtonian limit ($u^{t} \rightarrow 1$,
$u_{\varphi} \rightarrow \varpi^{2} \Omega$ where $\varpi$ is the
cylindrical radius of the star), this rotation law becomes
\begin{equation}
\Omega = \frac{A^{2} \Omega_{c}}{\varpi^{2} + A^{2}}.
\end{equation}
We use the technique of Ref.~\citep{SF95} to construct the rotating
equilibrium stars.  Our four differentially rotating equilibrium stars
that will be evolved using the full $\Gamma$-law equation of state of
Eq.~(\ref{gammalaw1}) are summarised in Table~\ref{tab:equilibrium}.

\subsection{Disk Mass from Equilibrium Configuration}
Before computing the collapse of the star, we can estimate the final
rest mass ratio between BH and its surrounding materials with the
technique of Ref.~\citep{SS02}.  The idea is the following.  Suppose a
star collapses axisymmetrically throughout the system.  In this case
the rest mass $m_{*}$ and the angular momentum $j_{*}$ along each
cylinder
\begin{eqnarray*}
m_{*} &=& \int_{V_{\rm cyl}} dv \sqrt{-g} \rho u^{t}
,\\
j_{*} &=& \int_{V_{\rm cyl}} dv \sqrt{-g} \rho u^{t} h u_{\varphi}
,
\end{eqnarray*}
are, in the absence of viscosity, conserved throughout the evolution.
Note that $V_{\rm cyl}$ is the three dimensional spatial volume along
the cylinder.  After the BH has formed, the material is swallowed at
least up to the radius of the innermost stable circular orbit (ISCO)
of the newly formed BH.  Ignoring material from outside the ISCO  we
can approximate the rest mass of the BH, the angular momentum of the
BH and the rest mass of the disk, defined as the mass outside the
apparent horizon of the newly formed BH, as
\begin{eqnarray}
m_{*}^{\rm BH} &=& \int_{V_{\rm isco}} dv \sqrt{-g} \rho u^{t}
,\\
j_{*}^{\rm BH} &=& 
  \int_{V_{\rm isco}} dv \sqrt{-g} \rho u^{t} h u_{\varphi}
,\\
m_{*}^{\rm disk} &=& \int_{V_{\rm whl} - V_{\rm isco}} dv \sqrt{-g} \rho u^{t}
,
\end{eqnarray}
where $V_{\rm isco}$ is the three dimensional spatial volume inside
the cylindrical radius of ISCO, $V_{\rm whl}$ the whole three
dimensional spatial volume, and $V_{\rm whl} - V_{\rm isco}$ is the
three dimensional spatial volume outside the ISCO cylindrical radius
in each hypersurface.

In practise, we compute the specific angular momentum at ISCO $j_{\rm
  isco}$.  Note that $j \equiv h u_{\varphi}$ is a monotonic growing
quantity as a function of cylindrical radius, and that $j_{\rm isco}$
is a function of the cylindrical rest mass $m_{*}$ and the
approximate, dimensionless Kerr parameter $j_{*} / m_{*}^{2}$.  If
there is a maximum in $j_{\rm isco}$ as a function of the cylindrical
rest mass (cylindrical radius), the quantities at the maximum
(cylindrical rest mass and specific angular momentum) correspond to
the quantity of the newly formed BH if the equilibrium star
collapses.  After the determination of the cylindrical rest mass of
the BH, we can compute that of the disk from the total rest mass of
the system.  We can also compute the approximate Kerr parameter from
the quantities at the maximum.  To simplify the calculation our
estimates of the ratio between disk mass and the BH mass and the Kerr
parameter are computed in conformally flat spacetime (using the
methods of e.g.\ Ref.~\citep{Saijo04}) and are summarised in
Table~\ref{tab:equilibrium}.

\begin{figure}
\centering
\includegraphics[keepaspectratio=true,width=8cm]{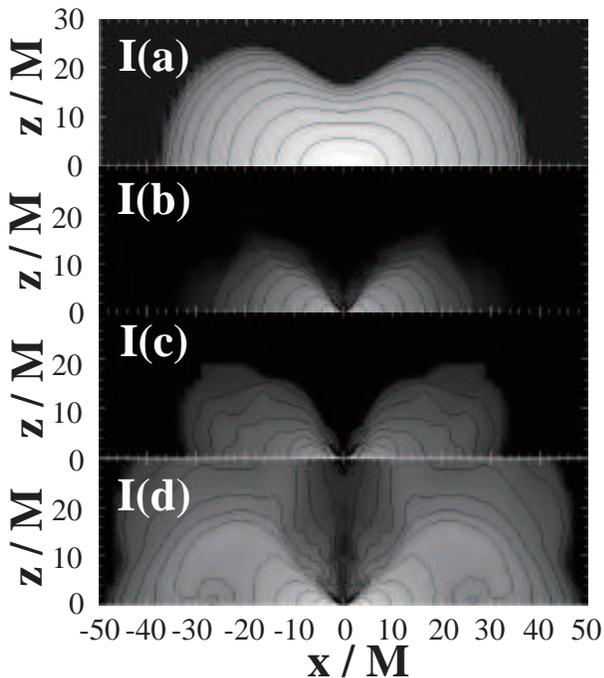}
\caption{
Same as Fig.~\ref{fig:qxy1}, but in the meridional plane. 
}
\label{fig:qxz1}
\end{figure}

\begin{figure*}
\centering
\includegraphics[keepaspectratio=true,width=14cm]{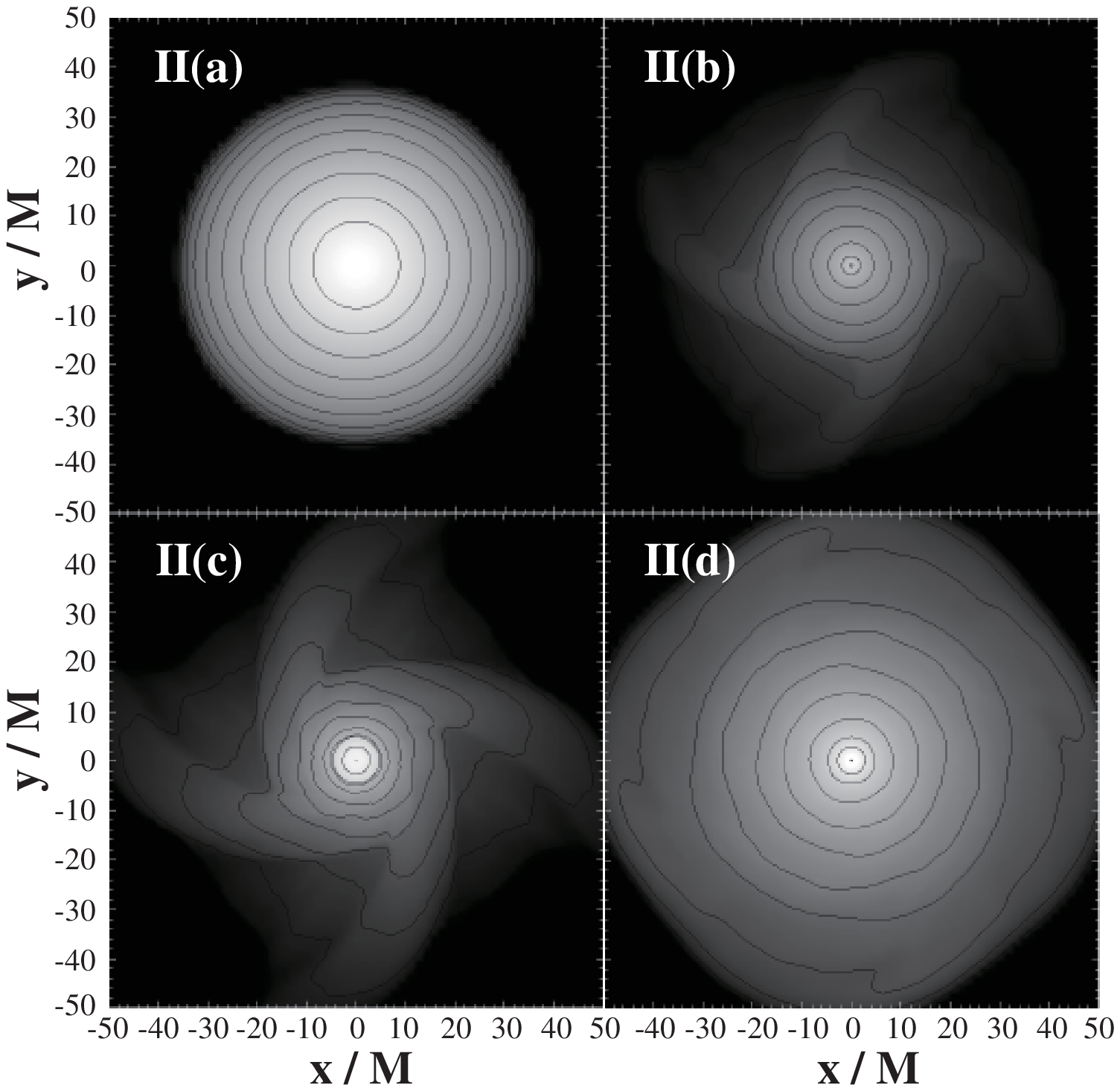}
\caption{
Contours of the rest mass density in the equatorial plane for model
II.  Snapshots are plotted at values of ($t$, $\rho_{\rm max}$) = (a)
($0$, $1.56 \times 10^{-5}$), (b) ($768 M$, $5.07 \times 10^{-2}$),
(c) ($849 M$, $1.05 \times 10^{-4}$) and (d) ($930 M$, $6.12 \times 
10^{-5}$), with a cutoff density of $\rho_{\rm cut} = 1.56 \times
10^{-13}$, respectively.   The contour lines denote rest mass
densities $\rho/\rho_{\rm max} = 10^{-(15-i)d}$ ($i=1, \cdots, 14$),
where $d=(\log \rho_{\rm max} - \log \rho_{\rm cut})/15$.  Note that
the apparent horizon exists after $t = 770 M$, and the coordinate
radius of the apparent horizon in the equatorial plane are $r_{\rm
  hrz} =$ (c) $0.170 M$ and (d) $0.175 M$, respectively. 
}
\label{fig:qxy2}
\end{figure*}

\begin{figure}
\centering
\includegraphics[keepaspectratio=true,width=8cm]{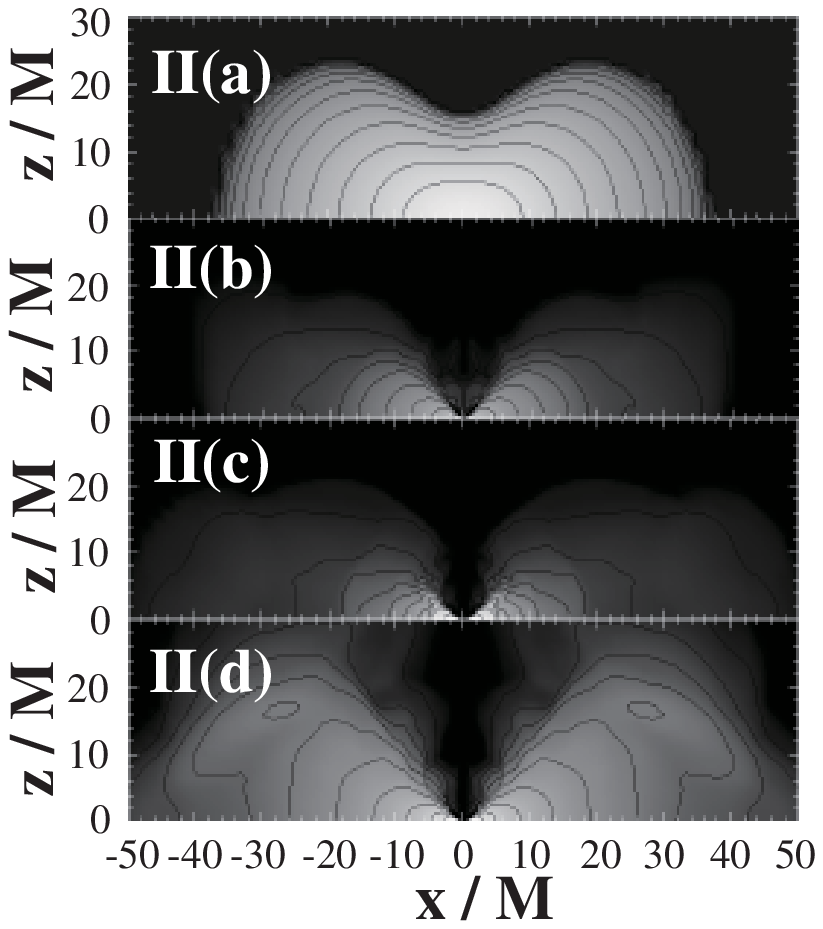}
\caption{
Same as Fig.~\ref{fig:qxy2}, but in the meridional plane. 
}
\label{fig:qxz2}
\end{figure}

\subsection{Collapse of Differentially Rotating Supermassive Stars}

We first investigate the onset of collapse by evolving the four
differentially rotating equilibrium stars of
Table~\ref{tab:equilibrium}.  We choose the $z$-axis as the rotational
one of the equilibrium star.  Since we use the polytropic equation of
state $P = \kappa \rho_0^{\Gamma}$ ($P$: pressure, $\kappa$: constant,
$\rho_0$: rest mass density, $\Gamma$: adiabatic exponent) when
constructing initial data sets, all physical quantities are rescalable
in terms of $\kappa$.  Therefore, we represent all physical quantities
nondimensionally in this paper, which is equivalent to setting
$\kappa=1$.  To trigger collapse, we deplete pressure by 1\%.
Checking the maximum rest mass density of the rotating stars
throughout the evolution, we conclude that models I and II are
radially unstable, while models III and IV are stable
(Fig.~\ref{fig:collapse}).

Since we use a free evolution scheme for the 17 variables associated
with the spacetime, the constraint equations of Einstein's field
equations are not automatically satisfied at each timestep.  Therefore
we can monitor the Hamiltonian constraint and Momentum constraints as
a check on the accuracy of the simulation.  We define them as 
\begin{eqnarray}
HC &=& R - K^{i}_{\;\;j} K^{j}_{\;\;i} + K^2 - 16 \pi \rho_{\rm H},\\
MC_i &=& \nabla_j K_i^{\;\;j} - \nabla_i K - 8 \pi S_i,
\end{eqnarray}
and show them for models I and II in Fig.~\ref{fig:const}.  Since
these quantities should be exactly zero sense, we normalise the
Hamiltonian constraint by $16 \pi \rho_{\rm max}$ and the Momentum
constraints by $8 \pi \rho_{\rm max} \sqrt{M/R}$, where $\rho_{\rm
  max}$ is the maximum rest mass density at each timestep.  Both of
the normalisation quantities represent the typical size of the source
term of the constraints (maximum rest-mass density and collapsing
momentum density).  Since all constraint quantities are conserved
within $\approx$1\% from Fig.~\ref{fig:const}, the constraints seem to
be satisfied at a reasonable level throughout the evolution.

We also monitor the rest mass $M_{0}$, ADM mass $M_{\rm adm}$, and the
total angular momentum $J$ 
\begin{eqnarray*}
M_{0} &=& \int dv \rho u^{t} \sqrt{-g},\\ 
M_{\rm adm} &=& - \frac{1}{2\pi} \oint_{\infty} \nabla^{i} \psi dS_{i} \\
 &= & \int [ (\rho h W^{2} -P) e^{5 \phi} + \frac{1}{16\pi}
    e^{5 \phi} A_{ij} A^{ij}]  ,\\
J &=& \frac{1}{8\pi} \oint_{\infty} 
  (x \tilde{A}_{yj} - y \tilde{A}_{xj}) e^{6 \phi} dS^{j} \\
&=& \int dv e^{6 \phi} [x S_{y} - y S_{x} + \frac{1}{8\pi} 
  (\tilde{A}^{x}_{y} - \tilde{A}^{y}_{x}) \\
&& - \frac{1}{16\pi} \tilde{A}_{ij} (x \partial_{y} - y
  {\partial}_{x}) \tilde{\gamma^{ij}} \\
&& + \frac{1}{12\pi}(x \partial_{y} - y
  {\partial}_{x}) K^{k}_{k}],
\end{eqnarray*}
throughout the evolution in Fig.~\ref{fig:raj}.  We only monitor these
three quantities until the time when the apparent horizon forms, since  
our excision method significantly complicates the computation of the
required volume integrals after horizon formation.  From
Fig.~\ref{fig:raj}, $M_{0}$ is conserved to within $0.5\%$, $M_{\rm
  adm}$ to within $1\%$, and $J$ to within $2\%$.  Therefore, we can
conclude that all three quantities are satisfactory conserved for our
case.

We also monitor the centre of mass condition in Fig.~\ref{fig:com}.
In practise we monitor
\begin{eqnarray}
M_x &=& \frac{\int dv^{\rm (half)} \rho x}{M_{0}^{\rm (half)}},\\
M_y &=& \frac{\int dv^{\rm (half)} \rho y}{M_{0}^{\rm (half)}},
\end{eqnarray}
where $M_{0}^{\rm (half)}$ is the rest mass of $z \geq 0$, $dv^{\rm
  (half)}$ the spatial volume of $z \geq 0$, since we adopt planar
symmetry across $z=0$.  Since our finest spatial gridsize is $\approx
3 \times 10^{-3} M$, the centre of mass motion is contained within a
few finest grids cells from the origin.  Therefore, we can conclude
that centre of mass condition is satisfied for our case.

We show  contours of the rest mass density in Figs.~\ref{fig:qxy1}
(equatorial plane) and \ref{fig:qxz1} (meridional plane) for model I
and in Figs.~\ref{fig:qxy2} (equatorial plane) and \ref{fig:qxz2} 
(meridional plane) for model II.  As we can easily understand from
these figures, the radially unstable stars first collapse to form a
BH.  After that, there is a significant amount of mass ejection due to
the large amount of angular momentum of the hole.  It is indeed like a 
flare.  Finally the system settles down to a quasi-stationary state;
central BH with a disk.  Fluid elements in the equilibrium star are
balanced by its self gravity, the pressure gradient and centrifugal
force.  When the BH forms, some fluid elements act as a particle which
is free from the interaction.  In this case, the material can spread
out to a larger radius than the equilibrium radius of the star.
Note that there seem to have $m=4$ structure in the snapshots of I (b)
(c) (Fig.~\ref{fig:qxy1}) and II (c) (Fig.~\ref{fig:qxy2}).  We
believe that this is due to the reduced spatial computational
resolution of the outer regions of the star (See Ref.~\citep{Saijo04}
for a different resolution case).

\begin{figure}
\centering
\includegraphics[keepaspectratio=true,width=8cm]{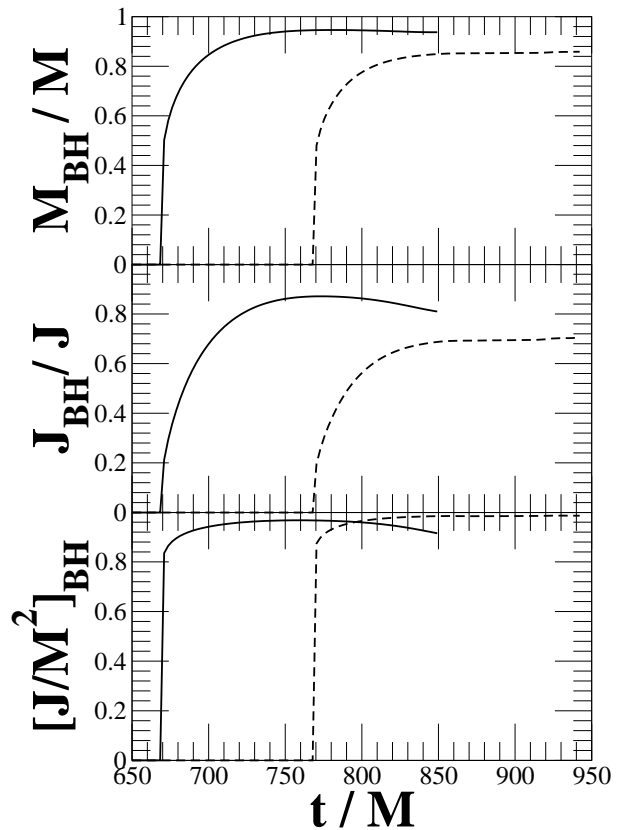}
\caption{
Gravitational mass ($M_{\rm BH}$), total angular momentum ($J_{\rm
  BH}$) and Kerr parameter ($(J/M^{2})_{\rm BH}$) of a newly formed BH
as a function of time.  Solid and dashed line represent models I and
II, respectively.
}
\label{fig:bh}
\end{figure}

Next we trace the mass and the angular momentum of the newly formed BH
throughout the evolution using the dynamical horizon techniques
outlined in Sec.~\ref{sec:DynamicalHorizon}.  We monitor the
gravitational mass, total angular momentum and the Kerr parameter of
the newly formed BH throughout the evolution as shown in
Fig.~\ref{fig:bh}.  The BH mass, the spin and the Kerr parameter
increase monotonically after the BH has formed, by swallowing much of
the surrounding material.  This stage lasts roughly until all of the
matter located inside the radius of the innermost stable circular
orbit of the final BH is swallowed.

\begin{figure}
\centering
\includegraphics[keepaspectratio=true,width=8cm]{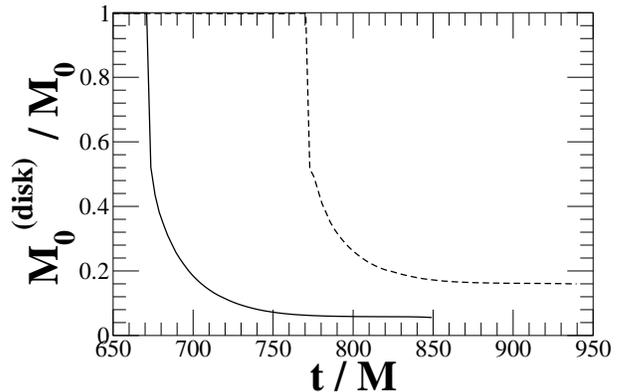}
\caption{
Disk mass as a function of time.  We defined the disk mass as the rest
mass outside the apparent horizon of the newly   formed BH.  Solid and
dashed line represent models I and II, respectively.
  }
\label{fig:disk}
\end{figure}

We have also confirmed that the estimated mass and spin of the BH from 
the equilibrium configuration of the collapsing SMS are in good
agreement with the results from the dynamics.   Although we calculated
the disk mass from the equilibrium configuration in conformally flat
spacetime, there is a little difference in the physical quantities
between full general relativistic spacetime and conformally flat
spacetime (see Table~\ref{tab:equilibrium}).  For instance, the
estimated Kerr parameter from the equilibrium star of model I is
$0.98$ (Table~\ref{tab:equilibrium}), while the result of numerical
simulation is $\approx 0.97$ (Fig.~\ref{fig:bh}).  Also the ratio
between the estimated rest mass of the disk and the rest mass of the
equilibrium star of model I is $0.044$ (Table~\ref{tab:equilibrium}),
while the result of numerical simulation is $\approx 0.05$
(Fig.~\ref{fig:disk}).  Note that we define the rest mass of the disk
as the rest mass outside the apparent horizon in each timeslice.

We furthermore study the formation of a massive disk from the collapse
of differentially rotating SMSs.  We trace the rest mass of the disk
for models I and II (Fig.~\ref{fig:disk}).   The rest mass of the disk
monotonically decreases once the BH has formed, since the newly formed
BH grows monotonically by swallowing the surrounding materials.  One
noticeable feature in Fig.~\ref{fig:disk} is that there is a plateau
at the final stage of model II.  This indicates that there is a strong
angular momentum barrier that a fluid fragments are prohibited to fall
into a hole.  We also illustrate the snapshot of the rest mass density
in the meridional plane of model II (Fig.~\ref{fig:qxz2} II[d]).  The
maximum of the rest mass density is located around $r \approx 2 M$ in
coordinate units.  

\begin{figure}
\centering
\includegraphics[keepaspectratio=true,width=8cm]{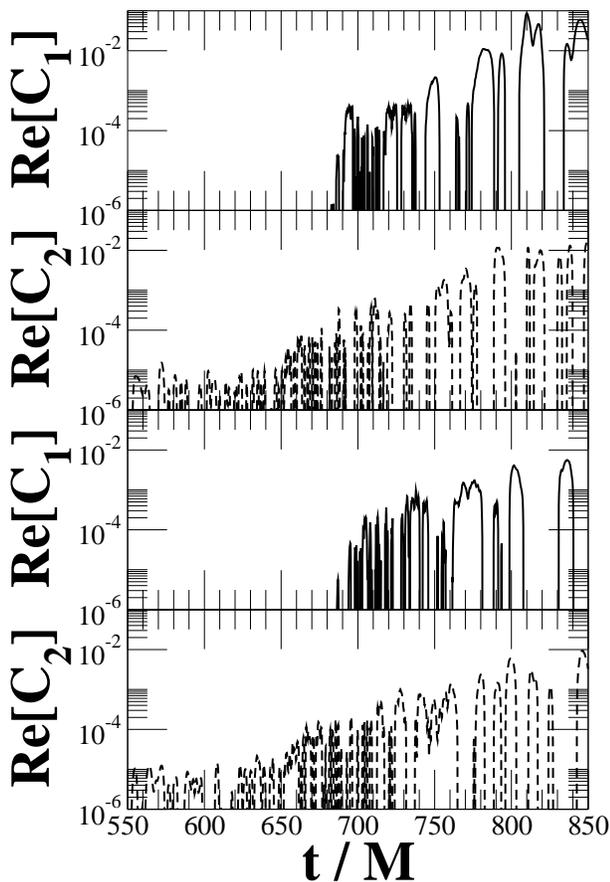}
\caption{
$m=1$ and $m=2$ diagnostics of the ring in the equatorial plane
throughout the collapse for model I.  We monitor the diagnostics at
the coordinate radius $r = 1.02 M$ (upper two panels) and $2.05 M$
(lower two panels), respectively.  Note that the solid line and the
dashed line denote  $m=1$ and $m=2$.  The coordinate equatorial radius
of the horizon at $t=854M$ is $r=0.464M$.
}
\label{fig:mode1a}
\end{figure}

\begin{figure}
\centering
\includegraphics[keepaspectratio=true,width=8cm]{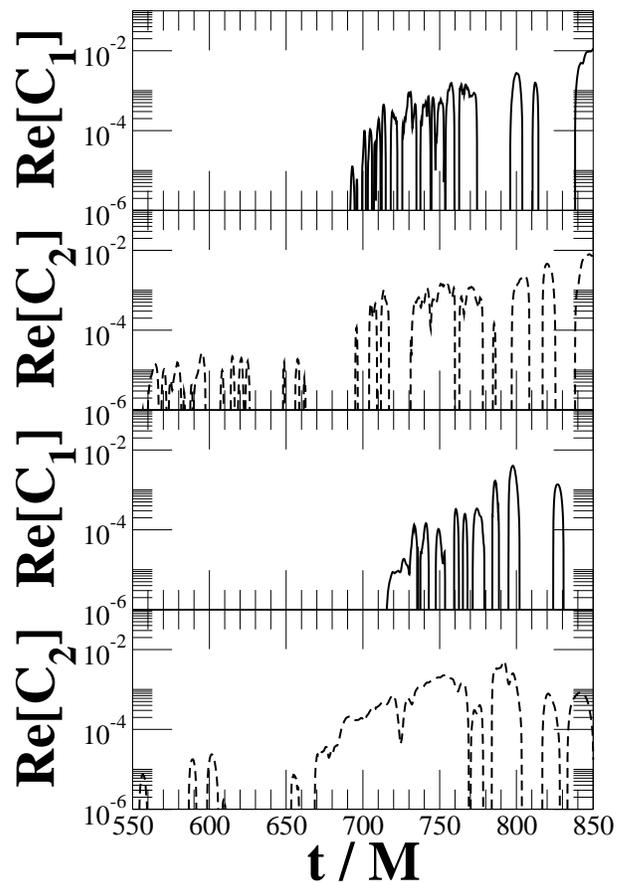}
\caption{
Same as Fig.~\ref{fig:mode1a}, but we monitor the diagnostics at the
coordinate radius $r = 4.1 M$ (upper two panels) and $10.2 M$ (lower
two panels), respectively.
}
\label{fig:mode1b}
\end{figure}

\begin{figure}
\centering
\includegraphics[keepaspectratio=true,width=8cm]{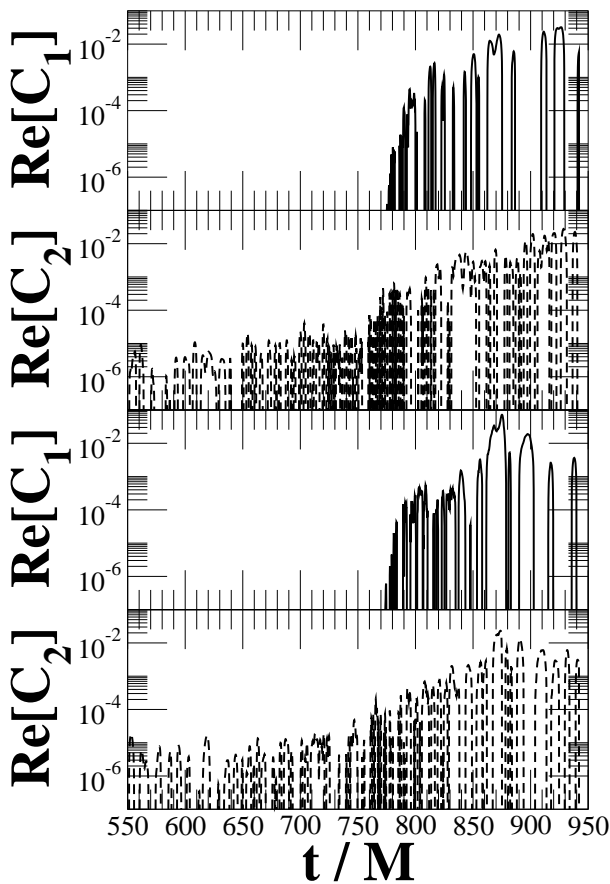}
\caption{
$m=1$ and $m=2$ diagnostics throughout the collapse for models II.  We
monitor the diagnostics at $r = 0.197 M$ (upper two panels) and $0.987
M$ (lower two panels), respectively.  Note that the solid line and the
dashed line denote $m=1$ and $m=2$.  The coordinate equatorial radius
of the horizon at $t=940M$ is $r=0.177M$.
}
\label{fig:mode2a}
\end{figure}

\begin{figure}
\centering
\includegraphics[keepaspectratio=true,width=8cm]{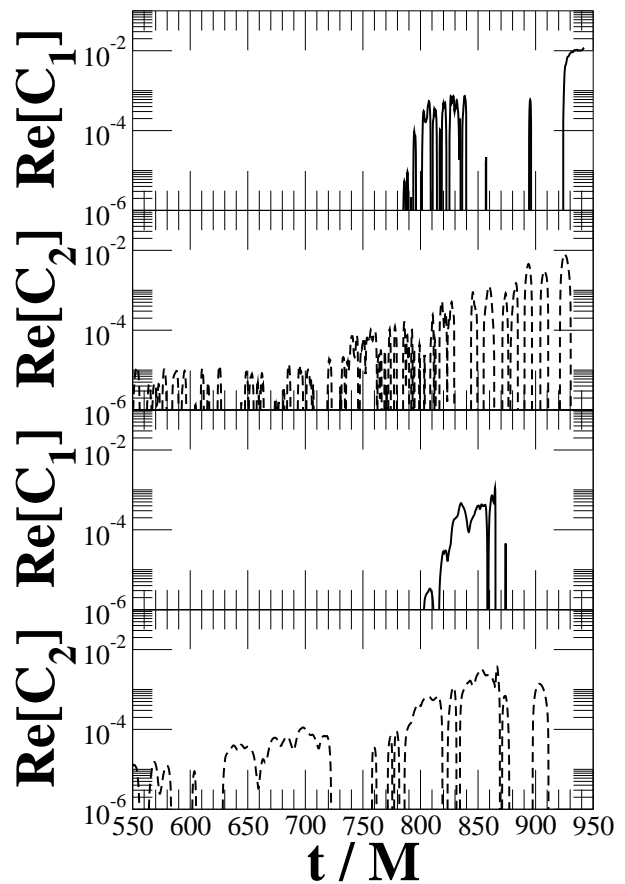}
\caption{
Same as Fig.~\ref{fig:mode2a}, but we monitor the diagnostics at $r =
3.95 M$ (upper two panels) and $9.87 M$ (lower two panels),
respectively.
}
\label{fig:mode2b}
\end{figure}

\begin{figure}
\centering
\includegraphics[keepaspectratio=true,width=8cm]{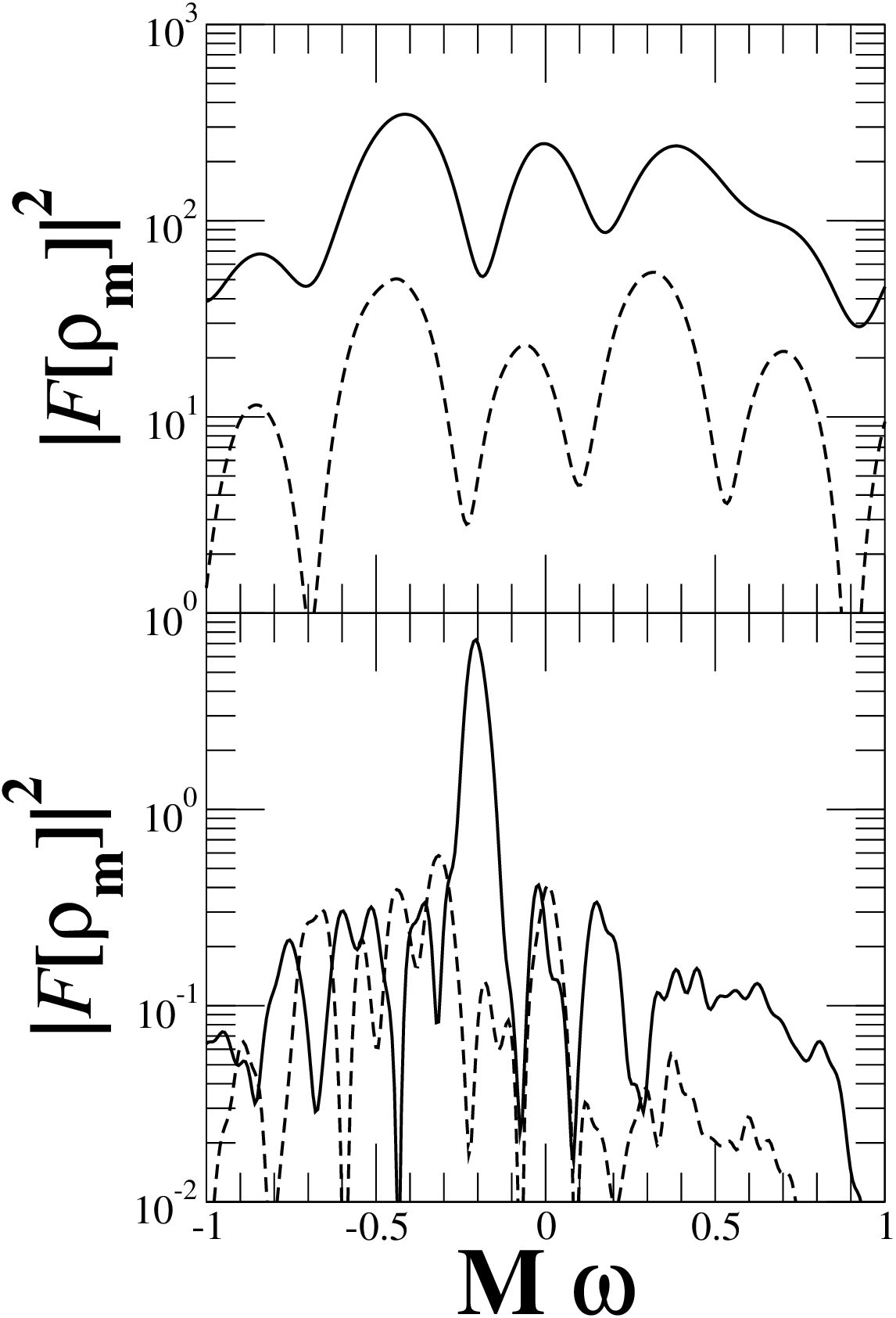}
\caption{
Spectra of $m=1$ and $m=2$ diagnostics throughout the collapse for
model I.  The top panel is monitored at the coordinate radius $r =
1.02 M$, while the bottom panel  $r = 2.05 M$.  Note that the solid
line and the dashed line denote $m=1$ and $m=2$.
}
\label{fig:modesp1}
\end{figure}

\begin{figure}
\centering
\includegraphics[keepaspectratio=true,width=8cm]{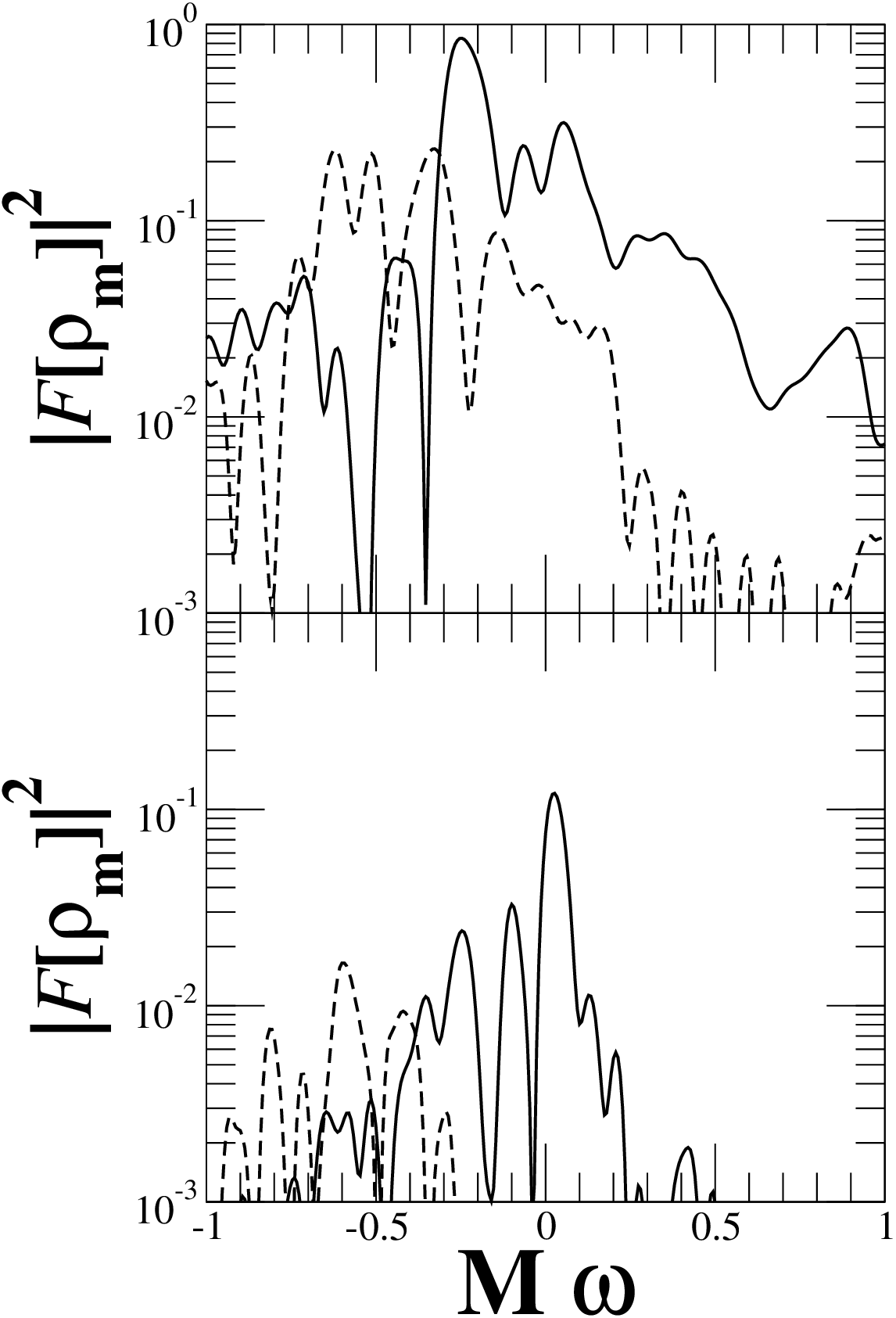}
\caption{
Spectra of $m=1$ and $m=2$ diagnostics throughout the collapse for
model II.   The top panel is monitored at the coordinate radius $r =
0.197 M$, while the bottom panel  $r = 0.987 M$.  Note that the solid
line and the dashed line denote $m=1$ and $m=2$.
}
\label{fig:modesp2}
\end{figure}

We also monitor the diagnostics of the $m=1$ and $m=2$ Fourier modes
throughout the collapse in Figs.~\ref{fig:mode1a} and \ref{fig:mode1b}
(for model I) and Figs.~\ref{fig:mode2a} and \ref{fig:mode2b} (for
model II).  We define the diagnostics $C_{m}$ as
\begin{equation}
C_{m} = \frac{1}{2 \pi} \int_{0}^{2\pi} \rho e^{i m \varphi} d\varphi,
\end{equation}
with a normalisation of $C_{0}$, a mean density of the ring.  In order
to understand the characteristic frequency, we also show the spectrum
of the ring diagnostics $|{\cal F}[\rho_{m}](\omega)|^{2}$ using the
Fourier transformation as
\begin{equation}
{\cal F}[\rho_{m}](\omega) = 
\frac{1}{4 \pi^2} \int_{0}^{t_{\rm fin}} dt \int_{0}^{2 \pi} \rho
  e^{i (\omega t - m \varphi)} d\varphi,
\end{equation}
where $t_{\rm fin}$ is the time where we terminate our simulation.
We show the spectra of the $m=1$ and $m=2$ diagnostics in
Figs.~\ref{fig:modesp1} and \ref{fig:modesp2}.  

The diagnostics show that non-axisymmetries (that would be triggered
at least at the level of finite-differencing error through the
simulation) do not grow to meaningful levels near the centre of the
system until well past horizon formation. When these non-axisymmetries
appear, they grow rapidly before saturating. From the growth of the
coefficients at larger radius, shown in Figs.~\ref{fig:mode1b} and
\ref{fig:mode2b}, it seems clear that the non-axisymmetries are
triggered in the interior of the disk and propagate outwards.

\subsection{Gravitational radiation}
\begin{figure}
\centering
\includegraphics[keepaspectratio=true,width=8cm]{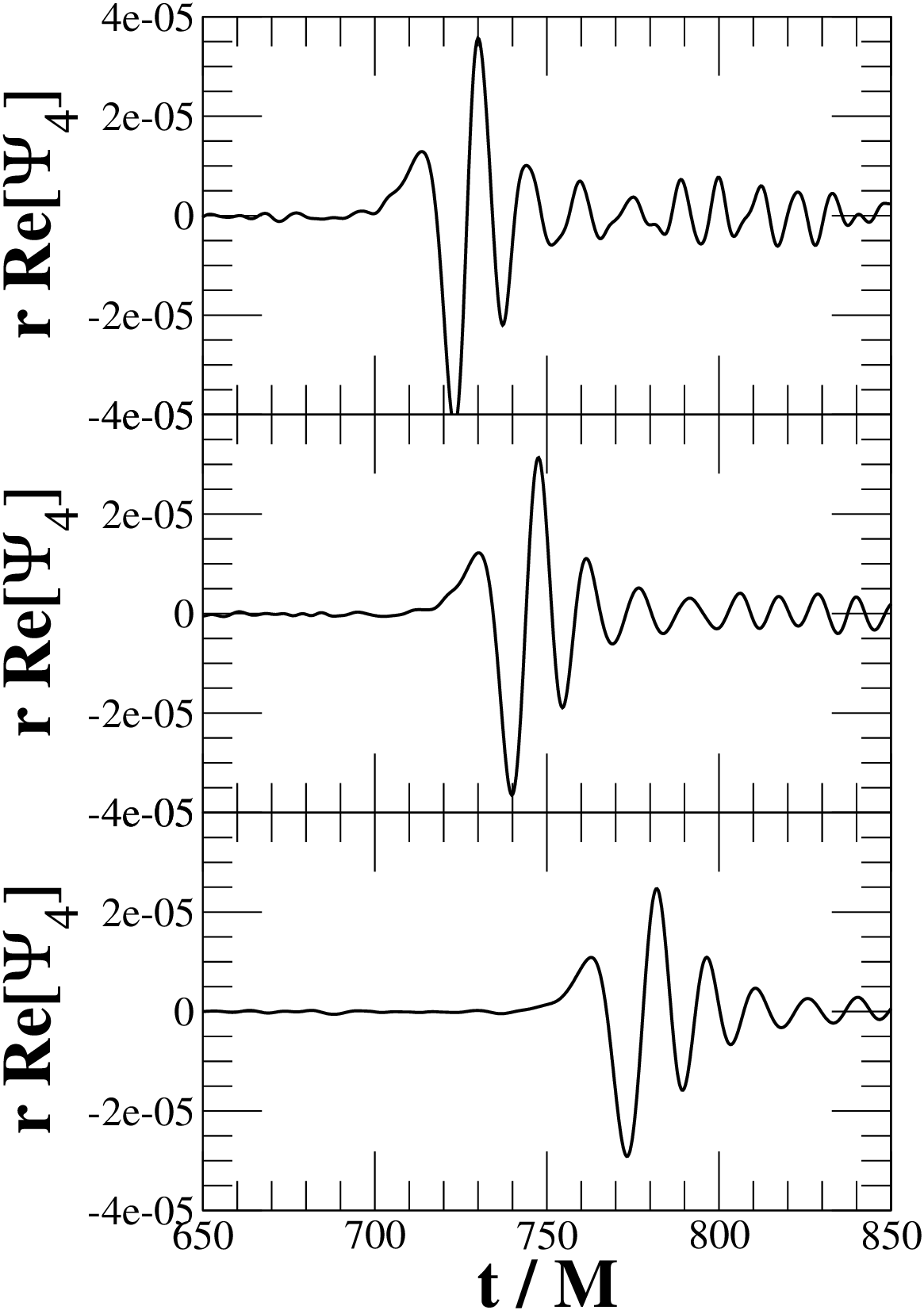}
\caption{
Gravitational waveform measured with the Weyl scalar $\Psi_4$ observed
along the $x$-axis in the equatorial plane at $r = 49.14 M$ (top
panel), $r = 65.52 M$ (middle panel) and $98.29 M$ (bottom panel) for
model I.  Note that the  time at which the apparent horizon is first
detected is $t=670M$.  Taking the wave propagation time from the
source to the observer into account, the apparent horizon formation in
the waveform is roughly just before the peak due to the burst.
Initially a standard burst and ringdown signal are seen
(compare~\citep{Buonanno07}, especially their Fig.~18), but sustained
gravitational wave emission indicate additional dynamics after BH
formation.
}
\label{fig:psi41}
\end{figure}

\begin{figure}
\centering
\includegraphics[keepaspectratio=true,width=8cm]{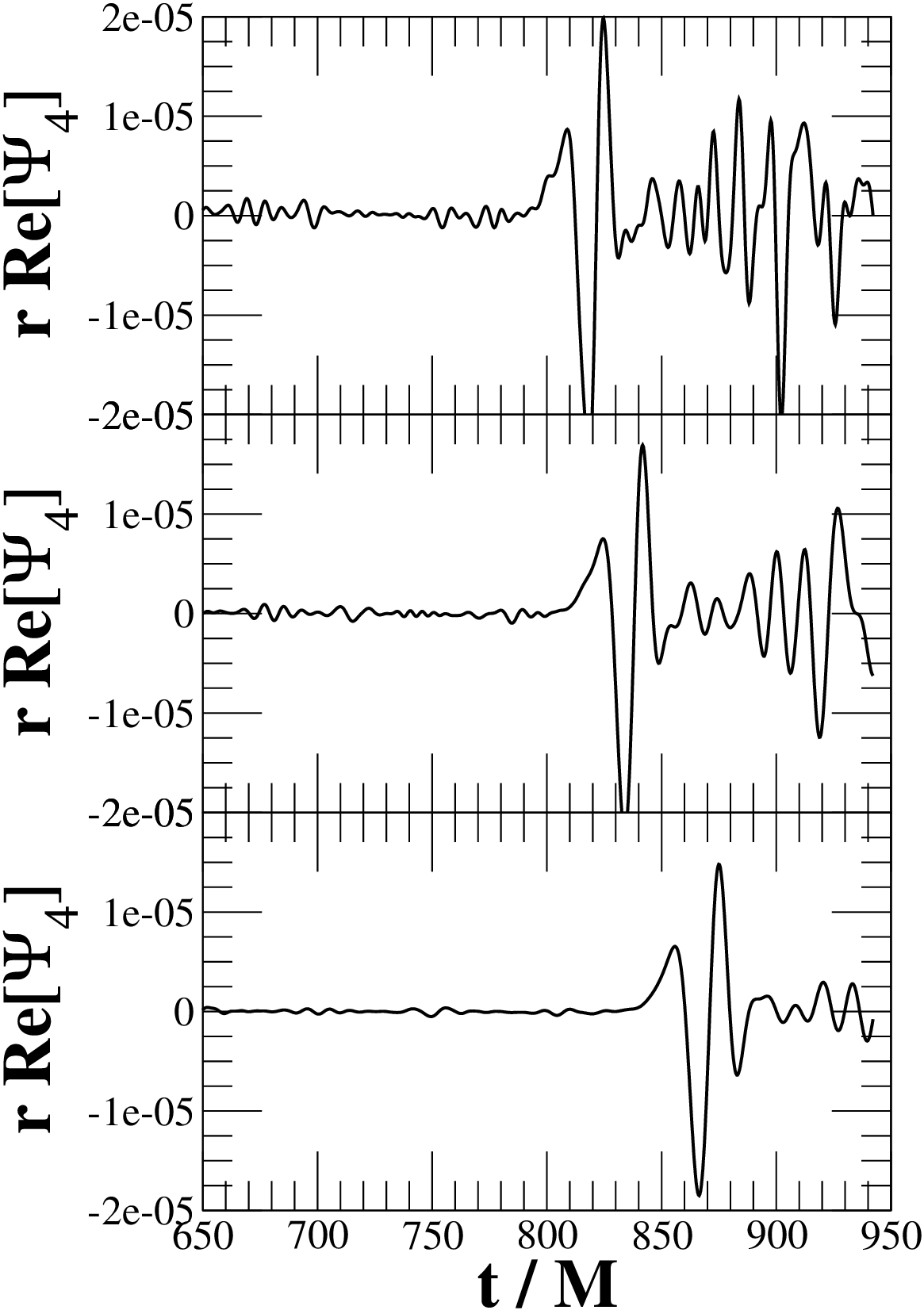}
\caption{
Gravitational waveform measured with the Weyl scalar $\Psi_4$ observed
along the $x$-axis in the equatorial plane at $r = 47.36 M$ (top
panel), $r = 63.15 M$ (middle panel) and $94.72 M$ (bottom panel) for
model II.  Note that the time at which the apparent horizon is first
detected is $t=770M$.  Taking the wave propagation time from the
source to the observer into account, the apparent horizon formation in
the waveform is roughly just before the peak due to the burst.
}
\label{fig:psi42}
\end{figure}

Finally we investigate the gravitational waveform from the collapsing
object.  We introduce the Weyl scalar $\Psi_4$ to study the outgoing
gravitational waves.  If we put the observer sufficiently far from the
source, the Weyl scalar $\Psi_4$ roughly represents the outgoing
gravitational waves, since the radiation back scattered by the
curvature behaves as $r^{-3}$, where $r$ is the distance from the
source.  We observe the waveform (the real component of $\Psi_4$)
along the $x$-axis in the equatorial plane at three different
coordinate locations with radius $r \approx 50 M$, $\approx 65 M$,
$\approx 100 M$ for models I and II (Figs.~\ref{fig:psi41} and
\ref{fig:psi42}).  Note that the equatorial coordinate radius of the
equilibrium star is $r = 38.1 M$ and $37.0 M$, and the outermost
boundary of the spatial grid is $131 M$ and $126 M$, for models I and
II.

\begin{figure}
\centering
\includegraphics[keepaspectratio=true,width=8cm]{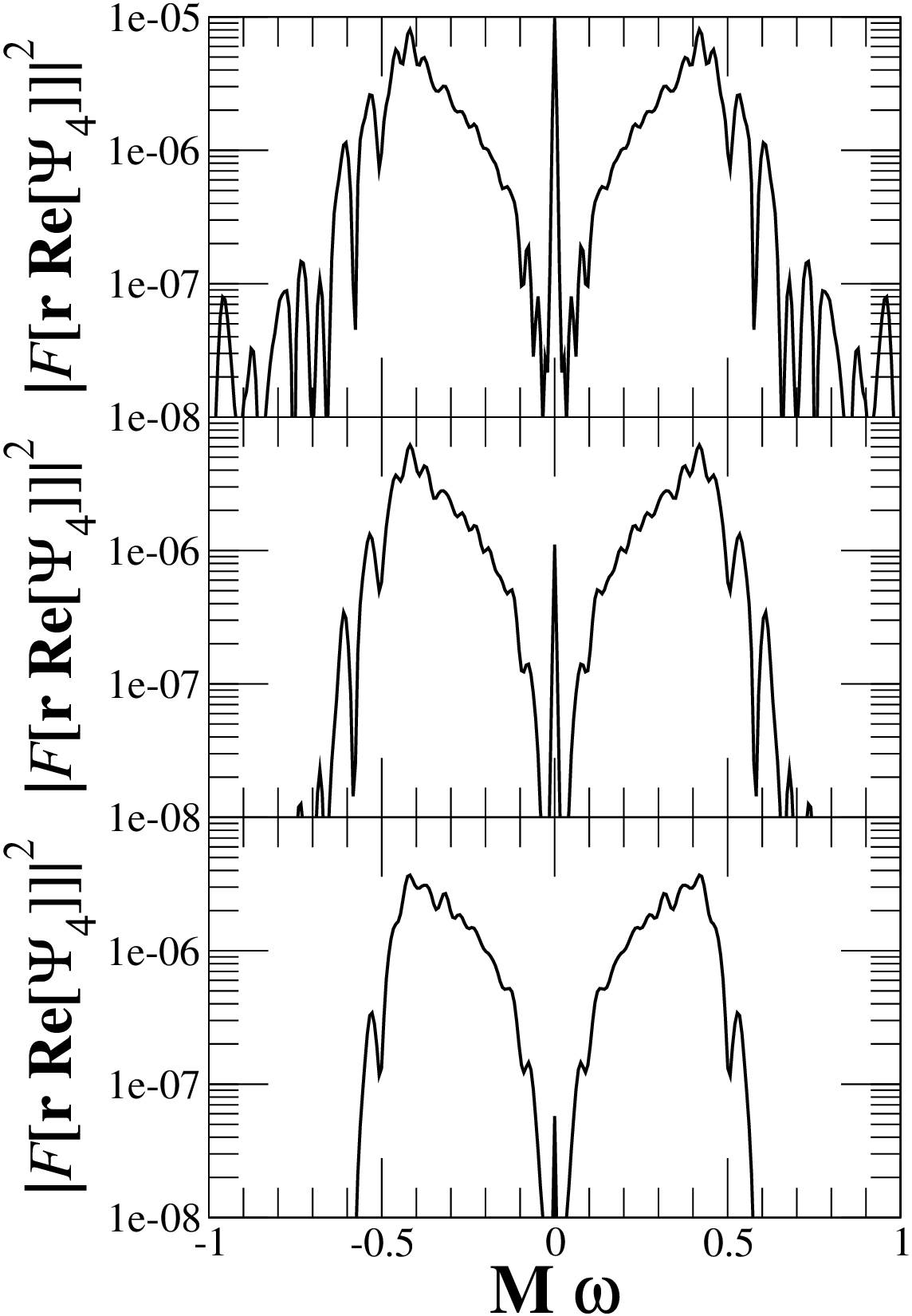}
\caption{
Spectra of the real part of the Weyl scalar $\Psi_4$ observed along
the $x$-axis in the equatorial plane at $r = 49.14 M$ (top panel), $r
= 65.52 M$ (middle panel) and $98.29 M$ (bottom panel) for model I.
}
\label{fig:sppsi41}
\end{figure}

We find that the waveform contains three different stages.  The first 
stage is the burst.  This occurs around the time when the apparent
horizon of the SMS forms.  The dominant contribution of the burst
comes from the axisymmetric mode due to collapse.  The second stage is
the quasinormal ringing of the newly formed BH.  Since the dominant 
frequency in the spectrum (Figs.~\ref{fig:sppsi41} and
\ref{fig:sppsi42}) is $M\omega \approx 0.4$, the dominant contribution
is the axisymmetric one, using the fact that quasinormal mode
frequency of $l=2$, $m=0$ is $M\omega \approx 0.40$ -- $0.42$ for a BH
with Kerr parameter $a \gtrsim 0.8 M$ \cite{Leaver85}, where $l$, $m$
denotes the indices of spin $-2$-weighted spheroidal harmonics.   The
final stage has a quasi-stationary wave.  The amplitude of the
waveform for model II seems to increase at late times.  The feature
corresponds to several peaks in the spectrum (Fig.~\ref{fig:sppsi42}),
especially in the regime of $M\omega \lesssim 0.5$.  For model I, the
quasi-stationary waveform remains for at least $\Delta t \sim 70M$
(Fig.~\ref{fig:sppsi42}).  In this case the spectra does not contain
the peaks in the regime of $M\omega \lesssim 0.5$.  When we reduce the
integration time of the quasi-periodic waves (comparing the three
panels in Fig.~\ref{fig:sppsi42}), the spectrum becomes much smoother
in the frequency regime $M\omega \gtrsim 0.5$.  Therefore the
frequency region of $M\omega \gtrsim 0.5$ plays a role in the
quasi-periodic waveform.  Although we do not know the origin of the
late-time quasi-periodic waves, we can discuss possible candidates and
their likelihood.

\begin{figure}
\centering
\includegraphics[keepaspectratio=true,width=8cm]{fig20.eps}
\caption{
Spectra of the Weyl scalar $\Psi_4$ observed along the $x$-axis in the
equatorial plane at $r = 47.36 M$ (top panel), $r = 63.15 M$ (middle
panel) and $94.72 M$ (bottom panel) for model II.
}
\label{fig:sppsi42}
\end{figure}

\begin{figure}
\centering
\includegraphics[keepaspectratio=true,width=8cm]{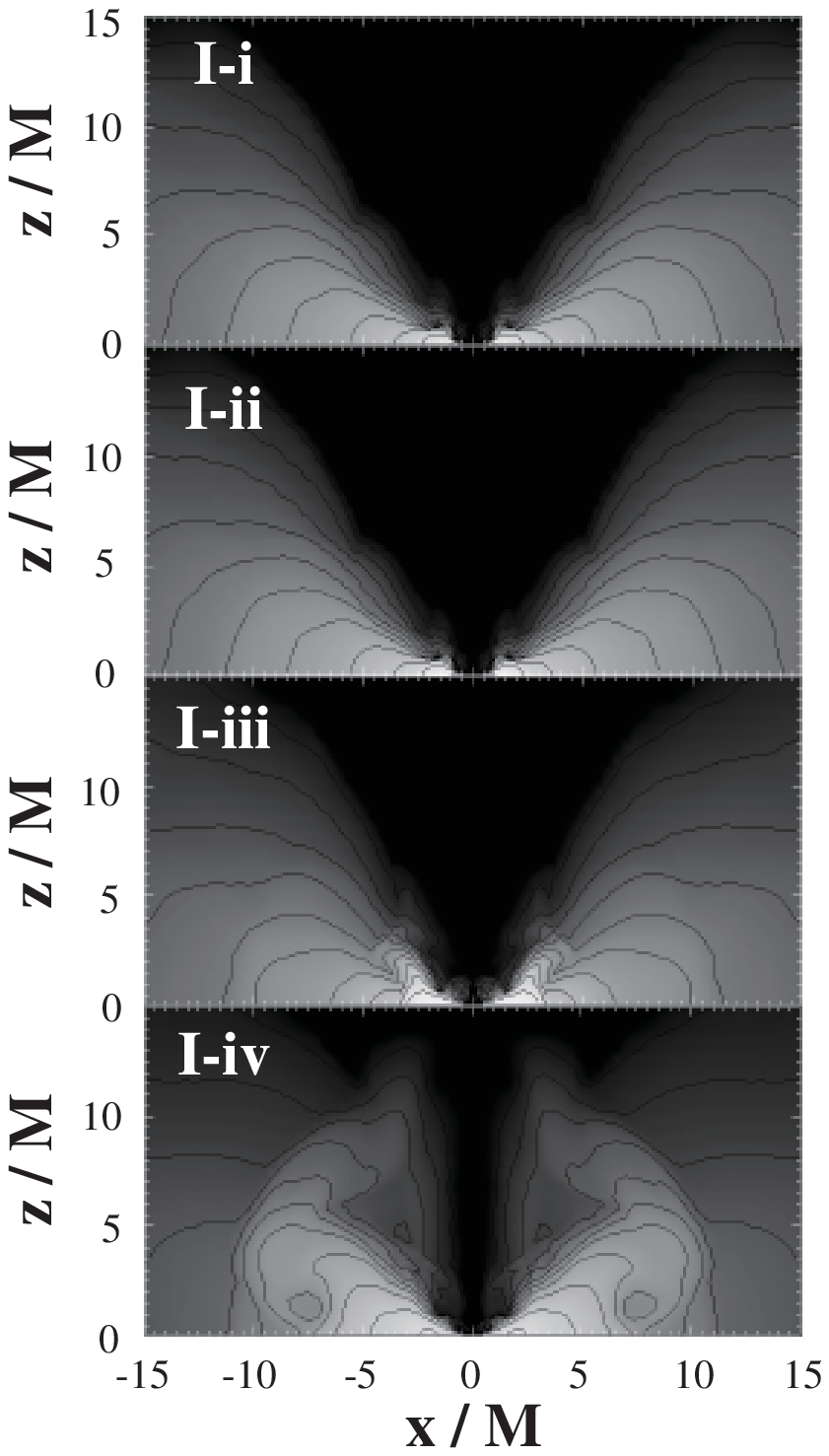}
\caption{
Contours of the rest mass density in the meridional plane for model
I.  Snapshots are plotted at values of ($t$, $\rho_{\rm max}$) = i
($713 M$, $1.87 \times 10^{-4}$), ii ($734 M$, $1.24 \times 10^{-4}$),
iii ($755 M$, $5.28 \times 10^{-5}$) and iv ($776 M$, $2.87 \times
10^{-5}$), with a cutoff density of $\rho_{\rm cut} = 1.56 \times
10^{-13}$, respectively.  The contour lines denote rest mass densities
$\rho/\rho_{\rm max} = 10^{-(15-i)d}$ ($i=1, \cdots, 14$), where
$d=(\log \rho_{\rm max} - \log \rho_{\rm cut})/15$.  Note that the
apparent horizon exists after $t = 671 M$, and the coordinate radius
of the apparent horizon in the equatorial plane are $r_{\rm hrz} =$
i $0.295 M$, ii $0.282 M$, iii $0.283 M$ and iv $0.297 M$,
respectively.  Shocks can be found in the snapshots of I-ii.
}
\label{fig:shock1}
\end{figure}

\begin{figure}
\centering
\includegraphics[keepaspectratio=true,width=8cm]{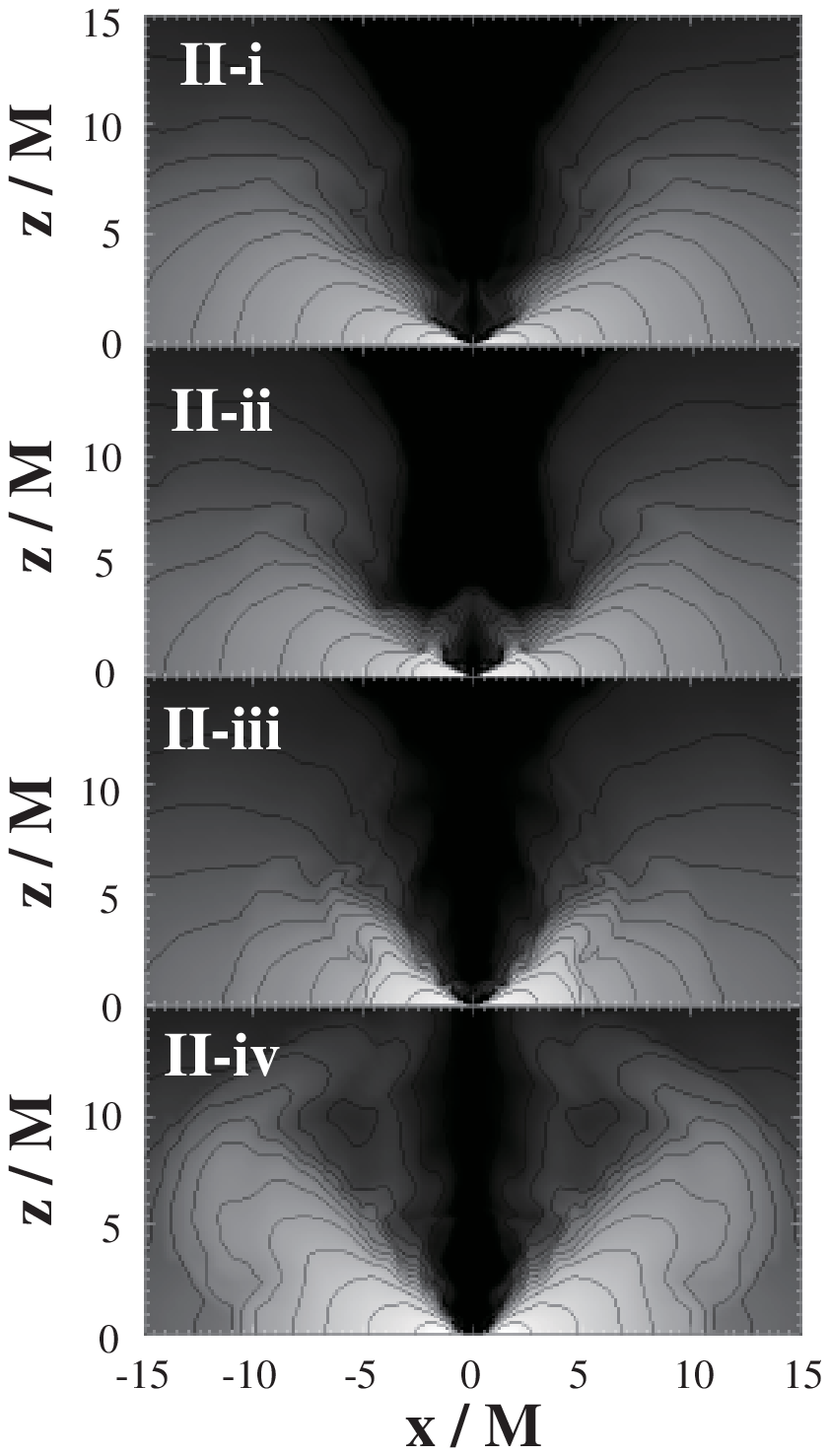}
\caption{
Contours of the rest mass density in the meridional plane for model
II.  Snapshots are plotted at values of ($t$, $\rho_{\rm max}$) = i
($808 M$, $3.15 \times 10^{-4}$), ii ($828 M$, $1.88 \times 10^{-4}$),
iii ($849 M$, $1.05 \times 10^{-4}$) and iv ($869 M$, $4.85 \times
10^{-5}$), with a cutoff density of $\rho_{\rm cut} = 1.56 \times
10^{-13}$, respectively.  The contour lines denote rest mass densities
$\rho/\rho_{\rm max} = 10^{-(15-i)d}$ ($i=1, \cdots, 14$), where
$d=(\log \rho_{\rm max} - \log \rho_{\rm cut})/15$.  Note that the
apparent horizon exists after $t = 770 M$, and the coordinate radius
of the apparent horizon in the equatorial plane are $r_{\rm hrz} =$
i $0.173 M$, ii $0.167 M$, iii $0.170 M$ and iv $0.180 M$,
respectively.  Shocks can be found in the snapshot of II-ii.
}
\label{fig:shock2}
\end{figure}

\begin{figure}
\centering
\includegraphics[keepaspectratio=true,width=8cm]{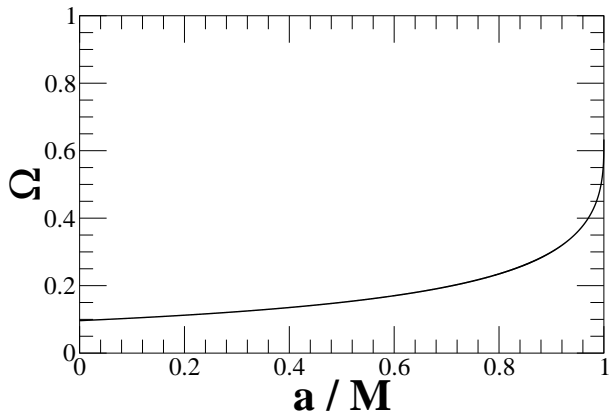}
\caption{
Angular velocity of circular orbit at ISCO radius as a function of
Kerr parameter.
}
\label{fig:isco}
\end{figure}

Firstly we distinguish between the different possibilities for the
amplification of the waves based on the dominant type of behaviour.
The most obvious possibility would be strong matter dynamics
perturbing the essentially stationary Kerr spacetime given by the
dominant central BH. The second possibility is a weaker, more
indirect coupling, where either the matter dynamics in the disk lead
to gravitational waves, or a global oscillation or resonance of whole
spacetime exists, including both the exterior matter and the BH.

In the first category where the wave emission is dominated by the
motion of the fluid in Kerr spacetime, there are two possibilities.
One is that inhomogeneities or radial motion in the disk exist,
leading to the wave emission. These can be viewed, to a first
approximation, as small fluid fragments behaving approximately like
test particles orbiting a Kerr BH. If we further approximate the
``particle'' to be moving in a circular orbit  then the characteristic
frequency $\omega_{\rm circular}$ and the amplitude $h_{\rm circular}$
from the quadrupole formula of gravitational waves can be written as 
\begin{eqnarray}
\omega_{\rm circular} &=& m \Omega_{r_0},\\
r h_{+~{\rm circular}} &=& 
  7 \times 10^{-1} \mu (M \omega)^{3/2} \cos \omega t,
\end{eqnarray}
where $\Omega_{r_0}$ is the orbital angular velocity of the particle
at the radius of ISCO $r_0$, and $r$ is the distance from the source.
The typical frequency for a circular orbit can be also applied for the
scattering particle at the innermost radius $r_0$.  We show the
relation between the orbital angular velocity at ISCO and the Kerr
parameter in Fig.~\ref{fig:isco}.  In fact, when the newly formed BH
is near extremal, the typical frequency $\omega_{\rm circular}$  and
the amplitude $h_{+~{\rm circular}}$ of gravitational waves are
\begin{eqnarray*}
M \omega_{\rm circular} &=& M \Omega_{r_0} \approx 0.5
,\\
r h_{+~{\rm circular}} &\approx& 2 \times 10^{-5} M 
  \left( \frac{M\omega}{0.5} \right)^{3/2},
\end{eqnarray*}
assuming that the only contribution to the amplitude comes from the
particle at ISCO (the ring mass is $\approx 3 \times 10^{-5}$), and
that $m=1$ is dominant ($\omega_{\rm circular} = \Omega_{r_0}$).  For
$m=1$ we have that $M \omega =M \Omega \approx 0.5$. This idea is 
suggested from the spectra of gravitational waveform
(Figs.~\ref{fig:sppsi41} and \ref{fig:sppsi42}).  At every observed
radius, there is a peak around $M \omega \approx 0.5$, which might
correspond to the motion of the particle at inner edge.  However this
idea cannot explain the enhancement of the quasi-periodic waves when
the newly formed BH is almost extremal. We also note that the Fourier
coefficients as shown in Figs.~\ref{fig:mode1a} -- \ref{fig:mode2b}
show non-axisymmetries growing in the interior and moving out far too
late (in retarded time) to cause the enhanced gravitational waves.

Remaining with the idea that the gravitational waves are generated by
the fluid motion, we note that the dynamics of the matter as shown by
Figs.~\ref{fig:qxy1} -- \ref{fig:qxz2} change after horizon
formation. Initially all matter collapses in. However, once the
quasi-stable disk forms in the interior there is still some infalling
matter left, which ``splashes'' onto the disk. This is reminiscent of 
the centrifugal bounce seen in early core-collapse simulations with a
simple equation of state (e.g.\ the Type II multiple-bounce models
of Ref.~\citep{Dimmelmeier02b}). As the rotation rate increases and the
spacetime is near extreme Kerr, the chances of shocks appearing
through some form of accretion are significantly increased, as shown
as far back as \citet{Wilson72}.  After submission, we note that
simulations of core collapse (\cite{MurphyOttBurrows09}) have seen
gravitational waves due to rapidly decelerated shocks. The shocks
formed by the accreting matter could lead to wave-packet like bursts
of gravitational waves. As these form in the outer layers of the disk
and predominantly propagate out, their effects are never seen in the
Fourier coefficients shown in Figs.~\ref{fig:mode1a} --
\ref{fig:mode2b}.  It is possible to approximately correlate the
shocks appearing in Figs.~\ref{fig:shock1} and \ref{fig:shock2} with
the start of the quasi-periodic waves in Figs.~\ref{fig:psi41} and 
\ref{fig:psi42}.  Rest-mass density snapshots in the meridional plane
show that shock waves are generated around $t \approx 734 M$ for model
I and $t \approx 828 M$ for model II.  Quasi-periodic waves in
$\Psi_4$ starts at $t \approx 780 M$ at $r \approx 50 M$ for model I,
at $t \approx 870 M$ at $r \approx 50 M$ for model II.  Since
radiation propagates at speed of light, those times for each model
roughly correspond with each other.  However, it seems unlikely that
these shocks could produce quasi-periodic waves of near constant
amplitude as seen in model I, where the newly formed BH is not an
extremal one (Fig.~\ref{fig:psi41}).

In the second category, we are looking for a property of the full
spacetime including both the BH and the matter. As the Kerr BH
contains the majority of the mass in the system, we would expect this
to dominate. So either there must be some special property of
quasi-normal modes of near extremal Kerr BHs, or there must be a
coupling through the spacetime between the matter in the disk and the
Kerr BH.  Since we know that the imaginary part of the quasi-normal
mode frequency goes to zero at extreme Kerr limit \cite{SN89}, there
are long-lived modes for near-extremal Kerr BHs, and hence
small perturbations from the exterior matter may produce long-lived
gravitational wave signals such as those seen in Fig.~\ref{fig:psi41}
for model I. It is not clear that this could easily lead to the strong
amplification seen for model II in Fig.~\ref{fig:psi42}. 

A final possibility that we consider in this category is a corotation
resonance.  In fact when the spin of the newly formed BH is very close
to extreme Kerr, the amplitude of the gravitational wave signal 
gradually grows after the quasinormal ringing.  We also check the
azimuthal $m$ modes of the rest mass density traced at the certain
radius in the equatorial plane, and found that the $m=2$ mode starts
growing exponentially after the ringdown (Fig.~\ref{fig:modesp2},
second top panel).  One possible explanation for the exponential
growth of the $m=2$ mode at late times is the existence of corotation
resonance of the newly formed disk triggered by the vibration of the
hole.   Note that the necessary condition to excite corotation
resonance in fluid mechanics is $\omega_{\rm chr} = m \Omega$
(e.g.\ Ref.~\citep{SY06}), where $\omega_{\rm chr}$ is the
characteristic frequency of the system,  $\Omega$ the angular
velocity.  If the corotation resonance is triggered by quasinormal
ringing, the necessary condition for triggering a corotation resonance
is $\omega_{\rm qnm} = m \Omega$ at a certain radius of the interior
star.  The real part of the quasi-normal mode frequency of $l=m=2$
increases with increasing Kerr parameter, and becomes $M\omega_{\rm
  qnm}= 1$ for the extreme Kerr BH.  The orbital velocity at ISCO also
increases with increasing Kerr parameter, and becomes $M\Omega_{r_0} =
(3-\sqrt{3})/2 = m\omega_{\rm particle}$ for the extreme Kerr BH
($\omega_{\rm qnm} < m \omega_{\rm particle}$).  Therefore, the
condition $\omega_{\rm qnm} = m \omega_{\rm paricle}$ is satisfied for
$m=2$ for a certain Kerr parameter near the extremal limit.

\section{Conclusion}
\label{sec:Conclusion}
We have investigated the collapse of differentially rotating SMSs,
especially focusing on the post BH formation stage, by means of three
dimensional hydrodynamic simulations in general relativity.  We
particularly focus on the onset of collapse to form a rapidly rotating
BH as a final outcome.

We have found that the qualitative results of the evolution for the
mass and spin of the final BH and disk are quite similar to
the estimates that can be computed from the equilibrium configuration
when the estimated, final BH has $J_{\rm (BH)}/M_{\rm (BH)}^{2} < 1$.
This result suggests that in the absence of a nonaxisymmetric
instability, the estimate of the BH mass and the disk mass agree
with a simple axisymmetric picture that the specific angular momentum
is conserved throughout the evolution, and the newly formed BH
swallows the matter up to the radius of the ISCO.

We have also found that a quasi-periodic wave occurs after the
ringdown of a newly formed BH.  As we would normally expect
the ringdown waveform to damp, it seems likely that the cause of this
waveform is due to the presence of the disk in some form. Furthermore,
when the newly formed BH is sufficiently close to extreme Kerr with
sufficient surrounding matter we have found that the wavesignal may be
significantly amplified.

We have discussed several possibilities for the origin of these
amplified waves. The most likely possibilities seem to be (a)
corotation resonance between the disk and the BH, (b) long-lived
gravitational waves from the near-extreme BH amplified by
perturbations in the disk, or (c) shocks from the infalling, accreting
matter.

It may be possible to distinguish between some of these possibilities
by further time integration, although at present the computational
expense is prohibitive. The wave-packet or burst nature of waves
generated by shocks, for example, should be possible to distinguish at
later times and at larger radii of extraction.  The slow power law
decay ($\sim 1/t$) in the extremal limit \citep{Andersson00} would,
however, be difficult to accurately capture at present with this type
of nonlinear numerical simulation.  Alternatives to free evolution
simulations such as the Fully Constrained Formalism of
Refs.~\citep{Bonazzola04,ICC08} should significantly improve the
accuracy and hence may be capable of capturing such effects, but the
specific gauge currently required by this formalism may make it
difficult to get a numerically well-behaved simulation of the extreme
cases investigated here.

Additional information about the horizon geometry could be found from
the multipole structure of the dynamical horizon (as defined in
Ref.~\cite{DHMultipole1}; various numerical implementations are
discussed in
Refs.~\cite{DH06,DHMultipole2,DHMultipole3,DHMultipole4}). In
particular this may indicate whether a strong perturbation of the BH
geometry is connected to the gravitational waves seen here. However,
it seems likely, comparing in particular to the results of
Refs.~\cite{DHMultipole3,DHMultipole4} where long term evolutions of
``clean'' binary black hole systems are considered, that better
numerical accuracy and longer integration times will be required in
order to rule out any but the strongest of perturbations.

Finally we discuss the detectability of gravitational waves from this
system.  The characteristic strength $h$ and the frequency $f$ of the
burst and ringing are (e.g. Ref.~\citep{SBSS02})
\begin{widetext}
\begin{eqnarray}
f_{\rm burst} &\sim& 
3 \times 10^{-2} 
\left( \frac{10^{6} M_{\odot}}{M} \right) 
\left( \frac{M}{R} \right)^{3/2}
[{\rm Hz}]
,\\
h_{\rm burst} &\sim& 
1 \times 10^{-18} \left( \frac{M}{10^{6} M_{\odot}} \right)
\left( \frac{1 {\rm G pc}}{d} \right)
\left( \frac{M}{R} \right)
,\\
f_{\rm QNM} &\sim&
2 \times 10^{-2} \left( \frac{10^{6} M_{\odot}}{M} \right) {\rm [Hz]}
, \\
h_{\rm QNM} &\sim&
6 \times 10^{-19} \left( \frac{\Delta E_{\rm GW}/M}{10^{-4}} \right)^{1/2}
\left( \frac{2 \times 10^{-2} {\rm [Hz]}}{f_{\rm QNM}} \right)^{1/2}
\left( \frac{M}{10^{6} M_{\odot}} \right)^{1/2}
\left( \frac{1 {\rm Gpc}}{d} \right)
,
\end{eqnarray}
\end{widetext}
where $d$ is the distance from the observer and $\Delta E_{\rm GW}$ is
the total radiated energy. We set $R/M = 1$, a characteristic mean
radius during BH formation.  Since the main targets of LISA are
gravitational radiation sources between $10^{-4}$ and $10^{-1}$ Hz, it
is possible that LISA can search for the burst and quasi-normal
ringing waves accompanying rotating SMS collapse and formation of an
SMBH.  Moreover, the frequency of the quasi-periodic waves after the
quasinormal ringing is quite similar to that of quasinormal ringing,
so is in the detectable regime.  Also the amplitude of the
quasi-periodic wave is roughly $10\%$ of the burst at the horizon, and
hence this feature can also be seen in LISA.

Although we focus on the case of the collapse of differentially
rotating SMS to a SMBH in this paper, our model is also applicable to
the collapse of population III stars which are also radiation pressure
dominated.  In this case, the mass range of the collapsing object
becomes of the order of a few hundred solar masses.  Therefore the most
sensitive region for the detection of burst waves and quasinormal
ringing waves becomes the order of hertz.  Such gravitational waves
might be seen in Deci-hertz interferometer gravitational wave
observatory (DECIGO) \citep{DECIGO}.

\acknowledgments
We would like to thank Nils Andersson, Leor Barack, Carsten Gundlach,
Tomohiro Harada, Akihiro Ishibashi, Michael Jasiulek, Kei-ichi Maeda,
Shinji Mukohyama and Erik Schnetter for discussions.  We would also
thank the anonymous referee for his/her careful reading of our paper.
MS furthermore thanks Eric Gourgoulhon and Misao Sasaki for their kind
hospitality at the Institute Henri Poincar\'{e} and at the Yukawa
Institute for Theoretical Physics, where part of this work was done.
This work was supported in part by the STFC rolling grant
(No.~PP/E001025/1), by the PPARC grant (No.~PPA/G/S/2002/00531) at the
University of Southampton, by the Special Fund for Research program
2008, 2009 in Rikkyo University, and by the Grant-in-Aid for the 21st
Century Centre of Excellence in Physics at Kyoto University.
Numerical computations were performed on the cluster in the Institute
of Theoretical Physics, Rikkyo University, on the Cray XT4 cluster in
the Centre for Computational Astrophysics, National Astronomical
Observatory of Japan, on the SGI-Altix3700 in the Yukawa Institute for
Theoretical Physics, Kyoto University, and on the myrinet nodes of
Iridis compute cluster at the University of Southampton.



\begin{thebibliography}{99}
%
\bibitem[\protect\citeauthoryear{Rees}{2003}]{Rees03}
M.~Rees, 
in {\it The Future of Theoretical Physics and Cosmology},
edited by G.~W.~Gibbons, E.~P.~S.~Shellard and S.~J.~Rankin 
(Cambridge Univ. Press, Cambridge, 2003), 17.
%
\bibitem[\protect\citeauthoryear{Baumgarte and Shapiro}{1999}]{BS99}
T.~W.~Baumgarte and S.~L.~Shapiro, 
\apj {\bf 526}, 941 (1999).
%
\bibitem[\protect\citeauthoryear{Saijo {\it et al.}}{2002}]{SBSS02}
M.~Saijo, T.~Baumgarte, S.~L.~Shapiro and M.~Shibata, 
\apj {\bf 569}, 349 (2002).
%
\bibitem[\protect\citeauthoryear{Shibata and Shapiro}{2002}]{SS02}
M.~Shibata and S.~L.~Shapiro, 
\apj {\bf 572}, L39 (2002).
%
\bibitem[\protect\citeauthoryear{Bodenheimer and Ostriker}{1973}]{BO73}
P.~Bodenheimer and J.~P.~Ostriker,
\apj {\bf 180}, 159 (1973).
%
\bibitem[\protect\citeauthoryear{Thorne}(1998)]{Thorne98}
K.~Thorne, 
in Black Holes and Relativistic Stars,
edited by  R. M. Wald (Univ. Chicago Press, Chicago), 41.
%
\bibitem[\protect\citeauthoryear{Sathyaprakash and Schutz}(2009)]{SS09}
B.~S.~Sathyaprakash and B.~F.~Schutz,
Living Rev. Relativity {\bf 12}, 2 (2009).
%
\bibitem[\protect\citeauthoryear{Fryer and New}(2003)]{FN03}
C.~L.~Fryer and K.~C.~B.~New,
Living Rev. Relativity {\bf 6}, 2 (2003).
%
\bibitem[\protect\citeauthoryear{New and Shapiro}{2001}]{NS01}
K.~C.~B.~New and S.~L.~Shapiro, 
\apj {\bf 548}, 439 (2001).
%
\bibitem[\protect\citeauthoryear{Saijo}{2004}]{Saijo04}
M.~Saijo, \apj {\bf 615}, 816 (2004).
%
\bibitem[\protect\citeauthoryear{Shibata and Nakamura}{1995}]{SN95}
M.~Shibata and T.~Nakamura,
\prd {\bf 52}, 5428 (1995).
%
\bibitem[\protect\citeauthoryear{Baumgarte and Shapiro}{1998}]{BS98}
T.~W.~Baumgarte and S.~L.~Shapiro,
\prd {\bf 59}, 024007 (1998).
%
\bibitem[\protect\citeauthoryear{Alcubierre {\it et al.}}{2003}]{ABDKPST03}
M.~Alcubierre, B.~Br\"ugmann, P.~Diener, M.~Koppitz, D.~Pollney,
E.~Seidel and R.~Takahashi,
\prd {\bf 67}, 084023 (2003).
%
\bibitem[\protect\citeauthoryear{Pollney {\it et al.}}{2007}]{PRRSADDDKNS07}
D.~Pollney, C.~Reisswig, L.~Rezzolla, B.~Szil\'agyi, M.~Ansorg,
B.~Deris, P.~Diener, E.~N.~Dorband, M.~Koppitz, A.~Nagar and
E.~Schnetter,
\prd {\bf 76}, 124002 (2007).
%
\bibitem[\protect\citeauthoryear{Baker {\it et al.}}{2006}]{BCCKM06}
J.~G.~Baker, J.~Centrella, D.-I.~Choi, M.~Koppitz and J.~van~Meter,
\prl {\bf 96}, 111102 (2006).
%
\bibitem[\protect\citeauthoryear{Banyuls {\it et al.}}{1997}]{BFIMM97}
F.~Banyuls, J.~A.~Font, J.~M.~Ib\`a\~nez, J.~M.~Mart\'i and
J.~A.~Miralles, 
\apj {\bf 476}, 221 (1997).
%
\bibitem[\protect\citeauthoryear{Collela and Woodward}{1984}]{Collela84}
P.~Collela and P.~R.~Woodward,
\jcp {\bf 54}, 174 (1984).
%
\bibitem[\protect\citeauthoryear{Aloy, Ib\`a\~nez, and Mart\'i}{1999}]{AIM99}
M.~A.~Aloy, J.~M.~Ib\`a\~nez and J.~M.~Mart\'i, 
\apjs {\bf 122}, 151 (1999).
%
\bibitem[\protect\citeauthoryear{Ashtekar and Krishnan}{2003}]{AK03}
A.~Ashtekar and B.~Krishnan, 
\prd {\bf 68}, 104030 (2003).
%
\bibitem[\protect\citeauthoryear{Schnetter, Krishnan and Beyer}{2006}]{DH06}
E.~Schnetter, B.~Krishnan and F.~Beyer, 
\prd {\bf 74}, 024028 (2006).
%
\bibitem[\protect\citeauthoryear{Dreyer {\it et al.}}{2003}]{DH03}
O.~Dreyer, B.~Krishnan, D.~Shoemaker and E.~Schnetter, 
\prd {\bf 67}, 024018 (2003).
%
\bibitem[\protect\citeauthoryear{Hawke, L\"offler and Nerozzi}{2005}]{Hawke05} 
I.~Hawke, F.~L\"offler and A.~Nerozzi, 
\prd {\bf 71}, 104006 (2005).
%
\bibitem[\protect\citeauthoryear{Baiotti and Rezzolla}{2006}]{BR06}
L.~Baiotti and L.~Rezzolla,
\prl {\bf 97}, 141101 (2006).
%
\bibitem[\protect\citeauthoryear{Goodale {\it et al.}}{2003}]{Cactus}
T.~Goodale, G.~Allen, G.~Lanfermann, J.~Mass\'o, T.~Radke,
E.~Seidel and J.~Shalf, 
Lecture Notes in Computer Science {\bf 2625}, 197 (2003);
\url{http://www.cactuscode.org}.
%
\bibitem[\protect\citeauthoryear{Schnetter, Hawley and Hawke}{2004}]{Carpet}
E.~Schnetter, S.~H.~Hawley and I.~Hawke, 
\cqg {\bf 21}, 1465 (2004);
\url{http://www.carpetcode.org}.
%
\bibitem[\protect\citeauthoryear{Baiotti {\it et al.}}{2005}]{Whisky1}
L.~Baiotti, I.~Hawke, P.~J.~Montero, F.~L\"offler, L.~Rezzolla,
N.~Stergioulas, J.~A.~Font and E.~Seidel, 
\prd {\bf 71}, 024035 (2005);
\url{http://www.whiskycode.org}
%
\bibitem[\protect\citeauthoryear{Baiotti, Giacomazzo and Rezzolla}{2008}]{WhiskyAccurate} 
L.~Baiotti, B.~Giacomazzo and L.~Rezzolla, 
\prd {\bf 78}, 084033 (2008). 
%
\bibitem[\protect\citeauthoryear{Stergioulas and Freidman}{1995}]{SF95}
N.~Stergioulas and J.~L.~Friedman,
\apj {\bf 444}, 306 (1995).
%
\bibitem[\protect\citeauthoryear{Leaver}{1985}]{Leaver85}
E.~W.~Leaver,
\prsla {\bf 402}, 285 (1985).
%
\bibitem[\protect\citeauthoryear{Buonanno, Cook and Pretorius}{2007}]{Buonanno07} 
A.~Buonanno, G.~B.~Cook and F.~Pretorius, 
\prd {\bf 75}, 124018 (2007). 
%
\bibitem[\protect\citeauthoryear{Dimmelmeier, Font and M\"uller}{2002}]{Dimmelmeier02b} 
H.~Dimmelmeier, J.~A.~Font, and E.~M\"uller, 
\aa {\bf 393}, 523 (2002).
%
\bibitem[\protect\citeauthoryear{Wilson}{1972}]{Wilson72}
J.~R.~Wilson,
\apj {\bf 173}, 431 (1972).
%
\bibitem[\protect\citeauthoryear{Murphy, Ott and Burrows}{2009}]{MurphyOttBurrows09}
J.~W.~Murphy, C.~D.~Ott and A.~Burrows, 
arXiv:0907.4762 (2009). 
%
\bibitem[\protect\citeauthoryear{Sasaki and Nakamura}{1989}]{SN89}
M.~Sasaki and T.~Nakamura, 
\grg {\bf 22}, 1351 (1989).
%
\bibitem[\protect\citeauthoryear{Saijo and Yoshida}{2006}]{SY06}
M.~Saijo and S.'i.~Yoshida, 
\mnras {\bf 368}, 1429 (2006).
%
\bibitem[\protect\citeauthoryear{Andersson and Glampedakis}{2000}]{Andersson00} 
N.~Andersson and K.~Glampedakis,
\prl {\bf 84}, 4537 (2000); 
K.~Glampedakis and N.~Andersson,
\prd {\bf 64}, 104021 (2001).
%
\bibitem[\protect\citeauthoryear{Bonazzola et al.}{2004}]{Bonazzola04} 
S.~Bonazzola, E.~Gourgoulhon, P.~Grandcl\'ement and J.~Novak, 
\prd {\bf 70}, 104007 (2004). 
%
\bibitem[\protect\citeauthoryear{Cordero-Carri\'on {\it et al.}}{2008}]{ICC08} 
I.~Cordero-Carri\'on, J.~M.~Ib\`a\~nez, E.~Gourgoulhon,
J.~L.~Jaramillo and J.~Novak,
\prd {\bf 77}, 084007 (2008). 

\bibitem[\protect\citeauthoryear{Ashtekar {\it et al.}}{2004}]{DHMultipole1}
A.~Ashtekar, J.~Engle, T.~Pawlowski and C.~Van Den Broeck, 
\cqg {\bf 21}, 2549 (2004).

\bibitem[\protect\citeauthoryear{Vasset {\it et al.}}{2009}]{DHMultipole2}
N.~Vasset, J.~Novak and J.~L.~Jaramillo,
\prd {\bf 79}, 124010 (2009). 

\bibitem[\protect\citeauthoryear{Jasiulek}{2009}]{DHMultipole3}
M.~Jasiulek, 
arXiv:0906.1228 (2009).

\bibitem[\protect\citeauthoryear{Owen}{2009}]{DHMultipole4}
R.~Owen, 
arXiv:0907.0280 (2009).

\bibitem[\protect\citeauthoryear{Kawamura {\it et al.}}{2006}]{DECIGO} 
S.~Kawamura {\it et al.}, 
\cqg {\bf 23}, S125 (2006).

\end{thebibliography}
\end{document}